\documentclass[aps,pra,reprint,amsmath,amssymb,longbibliography]{revtex4-2}
\usepackage[T1]{fontenc}
\usepackage{lmodern}
\usepackage{amsthm,bm,mathtools,graphicx}
\usepackage{microtype,capt-of,xcolor}
\usepackage{hyperref}
\hypersetup{breaklinks=true,pdfborder={0 0 0},colorlinks=true,linkcolor=blue,citecolor=magenta,urlcolor=violet,
 pdftitle={Path-Distinguishability Gaps at Fixed Visibility in Noisy Qubit Interferometers},
 pdfauthor={Mohd Asad Siddiqui}}
\newtheorem{theorem}{Theorem}
\newtheorem{proposition}{Proposition}
\newtheorem{corollary}{Corollary}

\newcommand{\Tr}{\operatorname{Tr}}

\newcommand{\norm}[1]{\left\lVert #1\right\rVert}
\newcommand{\bv}[1]{\bm{#1}}
\begin{document}
\title{Path-Distinguishability Gaps at Fixed Visibility in Noisy Qubit Interferometers}
\author{Mohd Asad Siddiqui}
\email{asad@ctp-jamia.res.in}
\affiliation{Department of Physics, Government Girls P.G. College, Sawai Madhopur, Rajasthan 322001, India}

\begin{abstract}
The wave--particle duality relation bounds visibility and path
distinguishability but does not determine their attainable values
in a fixed noisy interferometer. We characterize the complete
attainable region for a two-path interferometer with a qubit marker,
varying the marker input while fixing the source, preparation
channel, path-dependent interaction, and subsequent marker channel.
At each feasible fringe visibility, the Helstrom-optimal
distinguishabilities form at most two closed intervals, which can
be separated by a gap. The intervals arise from opposite axial
marker polarizations with the same fringe-amplitude modulus.
Fixing the complex fringe amplitude instead gives a connected
interval; for balanced original-path discrimination, a gap
requires nonunital marker preparation. For a balanced pure
source, relative $\pi$ rotations, and no subsequent marker noise,
we obtain a necessary and sufficient geometric criterion for a
gap under some rotation axis perpendicular to the preparation
displacement. Explicit channels and a time-independent Lindblad
realization show how the gap depends on preparation geometry and
evolution, with exact onset and optimization formulas at fixed
fractional visibility. We also give sufficient bounds for
excluding distinguishability intervals under calibrated
finite-resolution errors.
\end{abstract}

\maketitle
\raggedbottom

\section{Introduction}
\label{sec:intro}

Wave--particle complementarity connects interference visibility
to the available path information. Greenberger and Yasin quantified
a priori path predictability arising from unequal path populations
\citep{GreenbergerYasin1988}. Englert incorporated information
obtained from a which-way detector through optimal discrimination
of its conditional states, establishing the duality inequality
$V^2+D^2\leq1$ \citep{Englert1996}. This bound does not specify
which of the allowed pairs can actually be attained in a given
interferometer. Here we examine that question when the interactions
and noise channels are fixed and only the initial marker state
can be varied.

\begin{figure*}[t]
	\centering
	\includegraphics[width=\linewidth]{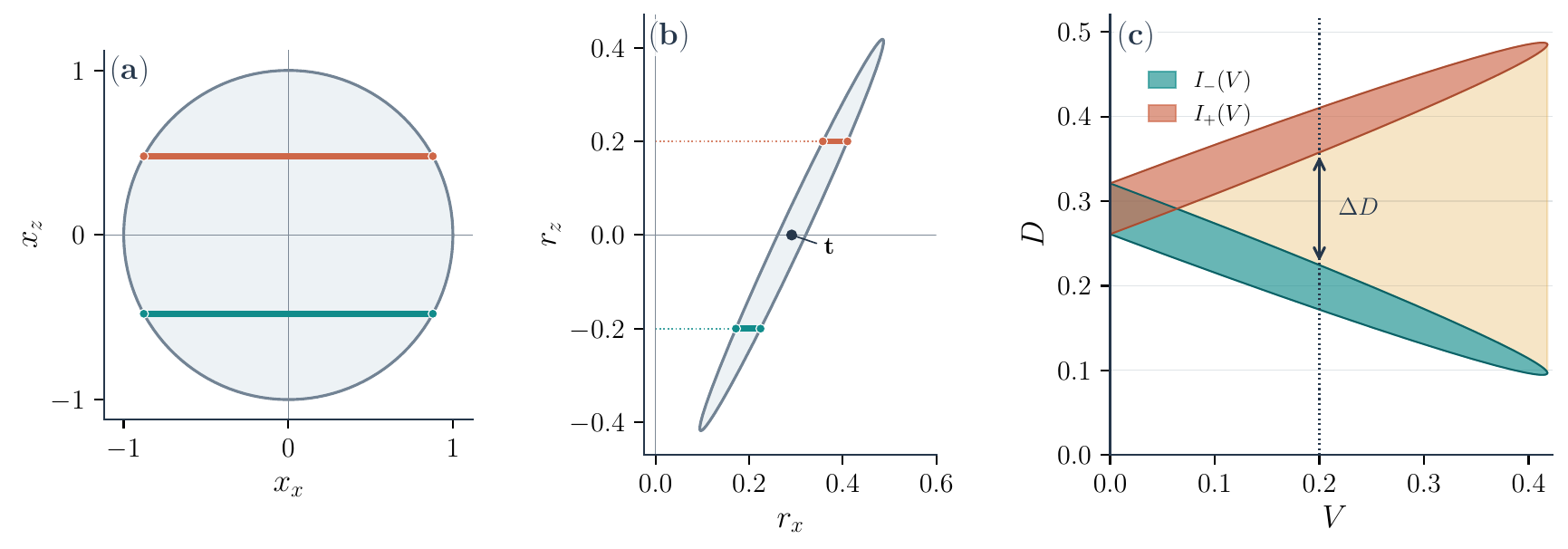}
	\caption{
		Geometry of a distinguishability gap at fixed visibility
		for the channel in Eq.~\eqref{eq:examplepars}, with a
		balanced pure quanton state, a relative $\pi$ rotation
		of the marker about $z$, and $\Psi=\mathrm{id}$.
		(a) The $x_y=0$ input cross section. At $V=0.2$,
		the planes $x_z=\pm V/b_0$ intersect the Bloch ball
		in two disks; the colored segments show their intersections
		with the plotted plane.
		(b) The corresponding prepared-marker sections in the
		$r_y=0$ plane lie at $r_z=\pm0.2$. The complete output
		disks are used to determine the distinguishability
		intervals.
		(c) Exact attainable bands $I_-(V)$ and $I_+(V)$,
		with the inaccessible interval shaded amber wherever
		the bands are separated. At the dotted slice,
		the gap width is $\Delta D\simeq0.1330$.
		The colored endpoints in panels (a) and (b)
		identify the corresponding boundary states.
	}
	\label{fig:gap}
\end{figure*}

In such an interferometer, the distinguishabilities attainable at
one visibility need not fill a single interval. A physical qubit
preparation channel constructed below produces two separated
ranges of Helstrom-optimal distinguishability, leaving intermediate
values inaccessible even though they satisfy the usual duality
inequality. Figure~\ref{fig:gap} illustrates the corresponding
marker-state sections. The gap concerns optimal path inference; it
neither violates complementarity nor excludes intermediate guessing
probabilities obtained from suboptimal measurements.

Wave--particle duality has been formulated using different
measures of path information and quantum coherence
\citep{Starke2026}. Siddiqui and Qureshi derived
a three-slit duality relation based on unambiguous state
discrimination \citep{SiddiquiQureshi2015}. Bera
\emph{et al.} developed a coherence-based formulation
for arbitrary multipath interference that recovers the
earlier three-slit relation \citep{Bera2015}. A later
work introduced a different multipath distinguishability
measure satisfying a quadratic coherence--distinguishability
relation \citep{QureshiSiddiqui2017}. Related approaches
consider a detector in a quantum superposition of presence
and absence \citep{SiddiquiQureshi2021}.
Bagan \emph{et al.} formulated multipath
coherence--path-information relations using minimum-error
state discrimination, recovering the binary Helstrom result
for two paths \citep{Bagan2016}, and later introduced
operational duality games \citep{Bagan2018}. These works establish discrimination-based measures
of path information, distinct from determining their attainable
range with restricted marker preparation.
L\"u obtained discrimination-based duality relations from path
asymmetry without an additional which-path detector
\citep{Lu2020}. The present setting instead uses conditional
marker states to infer the path.

Detector preparation and evolution also influence
interference. Banaszek \emph{et al.} showed how an internal
degree of freedom of the interfering particle affects the
relation between interference visibility and which-way
information \citep{Banaszek2013}. Sharma \emph{et al.}
studied how the initial states and conditional interactions
affect duality-bound saturation under detector noise
\citep{Sharma2020}. Their analysis concerns conditions for saturating the
duality bound; our question concerns attainability within that
bound.
Other work addresses state--channel coherence and
symmetry--asymmetry complementarity \citep{LuoSun2018},
path-selective decoherence \citep{MishraVenugopalanQureshi2019},
asymmetric-beam experiments \citep{Chen2022}, and entropic or
resource-theoretic descriptions \citep{Coles2014,Yang2024,Liu2025}.
These questions differ from the complete fixed-visibility
distinguishability set, which, to our knowledge, has not been
characterized for the noisy qubit-marker setting considered here.

The mechanism is geometric. Conventional visibility records
the modulus of a complex fringe amplitude, losing information
about its phase. Two opposite signs of a marker Bloch component
can therefore give the same visibility while selecting different
sections of the prepared-marker ellipsoid. Each section yields
an interval of distinguishabilities, and the intervals may fail
to overlap. Retaining the complex fringe amplitude removes
this ambiguity and gives a connected distinguishability range.

A geometric criterion singles out the role of the prepared-marker ellipsoid;
a pair of channels with the same displacement illustrates why
nonunitality alone is insufficient. Static channels and
Lindblad evolution realize the gaps explicitly. \mbox{Proofs of the endpoint formulas and} conditional finite-resolution bounds
are given in the \mbox{Supplemental Material}~\citep{SupplementalMaterial}.

\section{Fixed apparatus and operational quantities}
\label{sec:model}

We consider a two-path quantum system (quanton) $Q$ and a
qubit which-path marker (which-way detector) $M$, initially uncorrelated.
A fixed source prepares
\begin{equation}
	\begin{gathered}
		\rho_Q=
		\begin{pmatrix}
			p&z\\
			z^*&1-p
		\end{pmatrix},
		\qquad \alpha^2+\delta^2\leq1,\\
		\delta=2p-1,\quad
		P=|\delta|,\quad
		\alpha=2|z|=C_{\ell_1}(\rho_Q).
	\end{gathered}
	\label{eq:source}
\end{equation}
Here $P$ is the prior path predictability and
$C_{\ell_1}(\rho_Q)$ is the coherence in the original
path basis \citep{Baumgratz2014}. Predictability depends
only on the source populations, whereas path
distinguishability also includes information obtained
from the marker
\citep{GreenbergerYasin1988,Jaeger1995,
	MenonQureshi2018,Qureshi2021}.
The duality ellipse of Khatiwada and Qian relates visibility
to path predictability and the degree of coherence
\citep{KhatiwadaQian2025}. Its population-based predictability
corresponds to $P$ here; our attainable sections use the
marker-assisted distinguishability $D$ defined below.
Noise acting on the quanton before the marker interaction is absorbed
into $\rho_Q$.

The marker input ranges over the full Bloch ball,
$\norm{\bv x}\leq1$, where
$\norm{\bv x}=\sqrt{\bv x\cdot\bv x}$ is the Euclidean
norm. After the fixed preparation channel $\Phi$,
the marker state is
\begin{equation}
	\rho_\Phi=\tfrac12(\mathbb{I}_2+\bv r\cdot\bv\sigma),
	\qquad
	\bv r=T\bv x+\bv t.
	\label{eq:prep}
\end{equation}
Here $\mathbb{I}_2$ denotes the identity on the marker qubit;
$\mathbb{I}_3$ below denotes the identity on its real
three-dimensional Bloch-vector space.
The vector $\bv\sigma=(\sigma_x,\sigma_y,\sigma_z)$ denotes
the Pauli matrices, and $(T,\bv t)$ is the affine
representation of a completely positive trace-preserving
(CPTP) qubit channel \citep{Ruskai2002}. The displacement
$\bv t$ is the output Bloch vector for a maximally mixed
input, so $\Phi$ is unital if and only if $\bv t=0$.
Both pure and mixed marker inputs are allowed;
restricting the scan to pure states or a finite set
can change the attainable region.

The source, preparation channel, interaction, and subsequent
local channels remain fixed throughout the marker-input scan.

The path-dependent marker interaction is
$U=\sum_{j=0}^1|j\rangle\langle j|\otimes U_j$,
with associated Bloch-sphere rotations
$R_j\in SO(3)$. After the marker interaction, local channels
$\Lambda_Q$ on the quanton and $\Psi$ on the marker
give the joint state
\begin{equation}
	\Omega'=(\Lambda_Q\otimes\Psi)
	\big[U(\rho_Q\otimes\rho_\Phi)U^\dagger\big].
	\label{eq:jointstate}
\end{equation}
The post-interaction marker channel $\Psi$ is fixed and
path-independent. Its affine Bloch representation is
\begin{equation}
	\Psi:\ \bv y\longmapsto B\bv y+\bv h,
	\qquad
	\rho_j=\Psi(U_j\rho_\Phi U_j^\dagger).
	\label{eq:read}
\end{equation}
Here $B$ is a real $3\times3$ matrix and $\bv h$ is
its translation vector.
The baseline geometry takes $\Psi=\mathrm{id}$,
while the general results allow any fixed common
CPTP marker channel. Its environment is assumed
inaccessible.

The optimal probability of identifying the original
path, from a measurement on the marker, is
\begin{equation}
	P_{\mathrm{succ}}=\tfrac12(1+D),
	\qquad
	D=\norm{p\rho_0-(1-p)\rho_1}_1.
	\label{eq:Ddef}
\end{equation}
Here $\norm{X}_1=\Tr\sqrt{X^\dagger X}$ denotes
the trace norm. Consequently, $D\geq P$.
For balanced path priors, $p=1/2$, the
distinguishability reduces to the trace distance
between the conditional marker states. We call
this the \emph{original-path distinguishability},
to distinguish it from inference of the final path
after $\Lambda_Q$.
Abe and Edamatsu introduced a generalized distinguishability
for mixed markers that includes path information shared
with a purifying environment \citep{AbeEdamatsu2026}.
Equation~\eqref{eq:Ddef} instead quantifies optimal inference
from measurements on the marker alone.

Equation~\eqref{eq:Ddef} is the standard binary
minimum-error (Helstrom) distinguishability underlying
Englert's which-way relation and the two-path limit of
Bagan \emph{et al.}\ \citep{Helstrom1976,Englert1996,Bagan2016}.

After fixing the overall phase of the relative
marker unitary, we write
\begin{equation}
	\begin{gathered}
		W=U_1^\dagger U_0
		=\cos(\theta/2)\mathbb{I}_2
		-i\sin(\theta/2)\,\bv n\cdot\bv\sigma,\\
		0\leq\theta\leq\pi,\qquad \norm{\bv n}=1.
	\end{gathered}
	\label{eq:W}
\end{equation}
Here $\bv n$ is the relative rotation axis.
The prepared-marker Bloch vector has
parallel and perpendicular components
$r_\parallel=\bv n\cdot\bv r$ and
$\bv r_\perp=\bv r-r_\parallel\bv n$.

The two paths are recombined at an ideal balanced
beam splitter. The complex coherence factor and
conventional fringe visibility are
\begin{equation}
	\begin{aligned}
		\Gamma
		&=\Tr(\rho_\Phi W)\\
		&=\cos(\theta/2)
		-i\sin(\theta/2)r_\parallel,\\
		V
		&=\frac{I_{\max}-I_{\min}}{I_{\max}+I_{\min}}\\
		&=\alpha|\Gamma|\\
		&=\alpha\sqrt{\cos^2(\theta/2)
			+\sin^2(\theta/2)r_\parallel^2}.
	\end{aligned}
	\label{eq:V}
\end{equation}
Here $I_{\max}$ and $I_{\min}$ are the extrema of the
normalized intensity at one output port, obtained by scanning
the interferometric phase with balanced recombination.
To see how Eq.~\eqref{eq:V} follows, tracing out the marker
after the controlled interaction gives a reduced quanton
state with off-diagonal element $z\Gamma$. A relative
path phase $\varphi$ and balanced recombination then give
\begin{equation}
\begin{aligned}
 I(\varphi)
 &=\tfrac12\big[1+2\operatorname{Re}(z\Gamma e^{-i\varphi})\big]\\
 &=\tfrac12\big[1+\alpha|\Gamma|\cos(\varphi-\varphi_0)\big],
\end{aligned}
\label{eq:fringeintensity}
\end{equation}
where $\varphi_0=\arg(z\Gamma)$ when $z\Gamma\ne0$.
For $z\Gamma=0$, the intensity is independent of phase.
Thus $I_{\max,\min}=\tfrac12(1\pm\alpha|\Gamma|)$,
which gives the visibility in Eq.~\eqref{eq:V}.
The measured intensities are proportional to these
normalized detection probabilities.

For a pure prepared marker and before subsequent marker
noise, $\Gamma$ reduces to the overlap of the two
conditional marker states. For a mixed marker it is a
coherence factor, not the fidelity between those states.
For a pure quanton source,
$\alpha=\sqrt{1-\delta^2}$.
Immediately after the marker interaction, the $\ell_1$ coherence
of the reduced quanton state equals $V$ in this
recombination convention
\citep{Sharma2020,QureshiVisibility2019}.

The phase of $\Gamma$ shifts the fringes relative to the
source-coherence phase. With a calibrated phase reference,
the imaginary part of $\Gamma$ distinguishes the signs of
$r_\parallel$ when $\alpha\sin(\theta/2)>0$ and
$r_\parallel\neq0$. Our primary attainable region
uses phase-insensitive visibility; the phase-resolved
region is examined separately.

Throughout the main text, $V$ is evaluated after the marker
interaction and before $\Lambda_Q$, whereas $D$ refers to
inference of the original path from the marker after $\Psi$.
The common trace-preserving marker channel does not change
the reduced quanton state or its visibility. The optimal
positive operator-valued measure (POVM) may be chosen separately
for each input; equivalently, conditional-state tomography can
determine the optimum. Thus ``fixed apparatus'' specifies the
source and dynamics, while the readout is optimized for each
input. These quantities require direct access or calibrated
reconstruction at the stated stages. Supplemental
Sec.~\ref{sec:postquanton} treats observed visibility and
final-path inference under phase-covariant quanton noise.
Only local channels act after the marker interaction;
joint noise and postselection are excluded.

Define the weighted difference of conditional
Bloch rotations as $K_-=pR_0-(1-p)R_1$.
The qubit Helstrom formula then gives
\begin{equation}
	D=\max\{P,\norm{BK_-\bv r+\delta\bv h}\}.
	\label{eq:D}
\end{equation}
Indeed, the weighted difference of the conditional
states is $(\delta\mathbb{I}_2+\bv v\cdot\bv\sigma)/2$, with
$\bv v=BK_-\bv r+\delta\bv h$. Its eigenvalues are
$(\delta\pm\norm{\bv v})/2$, whose absolute sum
yields Eq.~\eqref{eq:D}. The translation $\bv h$ of the post-interaction
marker channel cancels for balanced priors but can contribute
when the priors are unequal.

The standard trace-distance and fidelity bounds imply
\begin{equation}
	V^2+D^2\leq1.
	\label{eq:standardduality}
\end{equation}
This is the established two-path duality relation
\citep{Englert1996,Bagan2016}, applied here to
original-path distinguishability and visibility
before $\Lambda_Q$. Its derivation is given in
Supplemental Sec.~\ref{supp:duality}.
To determine attainability, we identify all marker inputs
that realize a prescribed $V$ and calculate their range of $D$.

The minimum-error measure used here differs from
the unambiguous-discrimination quantity considered
in Refs.~\citep{SiddiquiQureshi2015,Bera2015}.
For nonzero path priors, a full-rank prepared marker
state gives zero unambiguous-discrimination success
probability after any common CPTP marker channel,
even when its Helstrom distinguishability is nonzero.
Further details are provided in Supplemental
Sec.~\ref{sec:usd}.

\section{Attainable regions and the role of fringe phase}
\label{sec:region}

We vary the marker input over the full Bloch ball,
including mixed states. For the operational quantities defined
in Sec.~\ref{sec:model}, let
\begin{equation}
	\mathcal R=
	\{(V(\bv x),D(\bv x)):\norm{\bv x}\leq1\}.
	\label{eq:R}
\end{equation}

The simplest case already reveals why a gap is possible.
Consider a balanced pure quanton, $\Psi=\mathrm{id}$,
and the path-dependent marker unitaries
$U_1=\mathbb{I}_2$ and $U_0=-i\,\bv n\cdot\bv\sigma$. Then
\begin{equation}
	V=|r_\parallel|,
	\qquad
	D=\norm{\bv r_\perp}.
	\label{eq:geometryVD}
\end{equation}
Indeed, the relative $\pi$ rotation sends the conditional marker
Bloch vector $\bv r$ to $2r_\parallel\bv n-\bv r$ on path $0$,
while path $1$ leaves it unchanged. For balanced paths, the
qubit trace-distance formula therefore gives
\[
\begin{gathered}
\bv r_0=2r_\parallel\bv n-\bv r,\qquad \bv r_1=\bv r,\\
D=\tfrac12\norm{\bv r_0-\bv r_1}=\norm{\bv r_\perp}.
\end{gathered}
\]
The longitudinal marker polarization sets the fringe contrast,
whereas its transverse polarization sets the optimal path information.
Fixing $V$ selects the planes
$\bv n\cdot\bv r=\pm V$ through the prepared-marker
Bloch ellipsoid, which may be degenerate. For each sign, the transverse distance $D$ ranges over an
interval. Although both sections have the same visibility,
their distinguishability ranges need not overlap: the usual
duality bound cannot see which signed section an input occupies.
The general calculation is simplest in the input Bloch ball,
where the two planar sections become disks. The example in
Sec.~\ref{sec:example} has two such sections at $V=0.2$,
with a strictly positive separation between their
Helstrom-distinguishability intervals.

\subsection{The two-disk characterization}

Express the distinguishability vector as an affine
function of the marker input and define
\begin{align}
	A&=BK_-T,&
	\bv d&=BK_-\bv t+\delta\bv h,
	\label{eq:Ad}\\
	\bv g&=T^{\mathsf T}\bv n,&
	\xi&=\bv n\cdot\bv t,\qquad
	G=\norm{\bv g}.
	\label{eq:g}
\end{align}
The vector $\bv g$ is normal to the visibility planes
in the input Bloch ball, while $\xi$ is the center
and $G$ the half-width of the axial projection.

Assume initially that $\alpha>0$, $\theta>0$, and
$G>0$. For $V\geq\alpha\cos(\theta/2)$, define
\begin{equation}
	u(V)=
	\frac{\sqrt{V^2/\alpha^2-\cos^2(\theta/2)}}
	{\sin(\theta/2)}.
	\label{eq:u}
\end{equation}
The visibility is feasible precisely when
\begin{equation}
	\max\{0,|\xi|-G\}
	\leq u(V)\leq|\xi|+G.
	\label{eq:udomain}
\end{equation}
Indeed, the axial component
$\bv n\cdot\bv r=\bv g\cdot\bv x+\xi$
ranges over $[\xi-G,\xi+G]$, and Eq.~\eqref{eq:V}
depends only on its absolute value. Thus fixing $V$ selects
up to two input disks, corresponding to $r_\parallel=\pm u(V)$;
only disks intersecting the Bloch ball contribute.

\begin{theorem}[Attainable fixed-visibility section]
	\label{thm:region}
	For the model of Sec.~\ref{sec:model}, assume
	$\alpha>0$, $\theta>0$, and $G>0$, and let $V$
	be feasible. For each $\sigma\in\{-1,+1\}$
	satisfying $|\ell_\sigma|\leq G$, define
	\begin{equation}
		\ell_\sigma=\sigma u(V)-\xi,\qquad
		\bv x_\sigma=\frac{\ell_\sigma}{G^2}\bv g,\qquad
		R_\sigma=\sqrt{1-\frac{\ell_\sigma^2}{G^2}}.
		\label{eq:disk}
	\end{equation}
	Let $E$ be a real $3\times2$ matrix whose orthonormal
	columns span the plane perpendicular to $\bv g$. Set
	\begin{equation}
		\begin{gathered}
			J=AE,\qquad
			\bv e_\sigma=A\bv x_\sigma+\bv d,\\
			f_\sigma(\bv y)=\norm{J\bv y+\bv e_\sigma}^2.
		\end{gathered}
		\label{eq:quad}
	\end{equation}
	Denote by $f_\sigma^-$ and $f_\sigma^+$ the
	minimum and maximum of $f_\sigma$ over
	$\norm{\bv y}\leq R_\sigma$. Then the attainable
	distinguishabilities at visibility $V$ are
	\begin{equation}
		\mathcal R_V=
		\bigcup_{\sigma:\,|\ell_\sigma|\leq G}
		\left[
		\max\{P,\sqrt{f_\sigma^-}\},
		\max\{P,\sqrt{f_\sigma^+}\}
		\right].
		\label{eq:section}
	\end{equation}
	Every value in each interval is attainable.
	If $\bv y_\sigma^\pm$ attains an extremum of
	$f_\sigma$, the corresponding marker input is
	$\bv x_\sigma+E\bv y_\sigma^\pm$.
	The two intervals must be retained separately
	unless they overlap.
\end{theorem}

\begin{proof}
	Equation~\eqref{eq:V} fixes
	$|\bv g\cdot\bv x+\xi|=u(V)$.
	For each feasible sign, the corresponding plane is
	$\bv g\cdot\bv x=\ell_\sigma$. Its intersection
	with the unit Bloch ball is
	\[
	\bv x=\bv x_\sigma+E\bv y,
	\qquad \norm{\bv y}\leq R_\sigma.
	\]
	Since $\bv x_\sigma$ is parallel to $\bv g$ and the
	columns of $E$ are orthonormal and perpendicular to $\bv g$,
	\[
	\norm{\bv x}^2=\frac{\ell_\sigma^2}{G^2}
	+\norm{\bv y}^2\leq1,
	\]
	which directly yields the radius $R_\sigma$ in
	Eq.~\eqref{eq:disk}. On this disk, Eq.~\eqref{eq:D} becomes
	$\max\{P,\sqrt{f_\sigma(\bv y)}\}$.
	A continuous real-valued function on a compact
	connected disk takes every value between its
	minimum and maximum. This establishes
	Eq.~\eqref{eq:section}, including collapsed
	intervals. The result does not depend on the
	choice of transverse basis $E$.
\end{proof}

The limiting cases require no singular extension
of the disk formulas. At $u=0$, the two signs
describe the same section. If $R_\sigma=0$,
the corresponding disk reduces to a single input.
For $G=0$, visibility is fixed at
\[
\alpha\sqrt{
	\cos^2(\theta/2)+
	\sin^2(\theta/2)\xi^2},
\]
and the entire input ball gives one
distinguishability interval. When $\alpha=0$,
visibility vanishes for every input and the same
full-ball optimization applies. For $\theta=0$,
the conditional marker states coincide and the
attainable region reduces to the point $(\alpha,P)$.

If the allowed inputs are instead restricted to pure states,
each feasible signed disk becomes a circle or point (or the
whole Bloch sphere when visibility is input independent).
These sets remain connected, so each distinguishability image
is still an interval, although its endpoints can differ from
those obtained by allowing mixed states.

Scalar endpoint equations and attaining inputs, including
singular quadratic cases, are given in
Supplemental Sec.~\ref{app:trust}.

\subsection{Phase-resolved sections and control constraints}
\label{sec:phase}

Conventional fringe visibility retains only the modulus of the
complex coherence factor. Using the source-coherence
phase as a calibrated reference, define
\begin{equation}
	\begin{aligned}
		\mathcal A(\bv x)
		&=\alpha\Gamma(\bv x)\\
		&=\alpha\left[
		\cos\frac{\theta}{2}
		-i\sin\frac{\theta}{2}
		(\bv g\cdot\bv x+\xi)
		\right],
	\end{aligned}
	\label{eq:complexamplitude}
\end{equation}
so that $V=|\mathcal A|$.
The imaginary part distinguishes the two signs of
the axial marker component that visibility
does not resolve.

\begin{theorem}[Connected sections at fixed complex fringe amplitude]
	\label{thm:phase}
	For the apparatus of Sec.~\ref{sec:model}, with
	arbitrary fixed path priors and any common CPTP
	post-interaction marker channel, the attainable
	distinguishabilities at a fixed feasible complex
	amplitude $\mathcal A=\mathcal A_0$ form a single closed interval.
	A disconnected fixed-visibility section can therefore
	arise only when taking the modulus combines distinct
	phase-resolved sections.
\end{theorem}

\begin{proof}
Fixing the complex amplitude imposes an affine constraint on the
input Bloch vector. Its feasible set is a disk, a point, or the
whole ball when the amplitude is input independent. Each set is
compact and connected, so the continuous distinguishability has
a closed interval as its image. Supplemental
Sec.~\ref{sec:phaseproof} gives the degenerate cases explicitly.
\end{proof}

The phase-resolved connectedness result also holds for finite-dimensional
markers with a compact convex set of allowed input states: a fixed complex
amplitude selects a convex subset whose continuous distinguishability image
is an interval (Supplemental Sec.~\ref{sec:phaseproof}).
The two-interval bound at fixed visibility, by contrast, relies on
qubit geometry and does not follow in arbitrary dimensions.

The complete region $\mathcal R$ is connected,
being the continuous image of the Bloch ball,
even though its fixed-visibility sections can
be disconnected [Fig.~\ref{fig:gap}(c)].
The gap is therefore a property of the fixed-visibility
slice, not a separation of the entire attainable region.

The two-branch structure has a further consequence when
marker inputs are adjusted continuously, rather than switched
discontinuously between settings.

\begin{corollary}[Minimum visibility during continuous branch switching]
	\label{cor:switch}
	Suppose $\alpha\sin(\theta/2)>0$ and a feasible
	visibility $V_0>\alpha\cos(\theta/2)$ has two
	disjoint distinguishability intervals.
	Any continuous marker-input trajectory connecting
	an operating point in one interval to a point in
	the other must pass through
	\begin{equation}
		V_{\min}=\alpha\cos(\theta/2).
		\label{eq:visibilitycost}
	\end{equation}
	The visibility must therefore decrease by at least
	$V_0-V_{\min}$ during the switching process.
	For a relative $\pi$ rotation, it must vanish at
	an intermediate input.
\end{corollary}

\begin{proof}
	The two branches have opposite axial components,
	$\bv n\cdot\bv r=\pm u(V_0)$. Every continuous
	trajectory connecting them crosses zero axial component,
	where Eq.~\eqref{eq:V} gives $V=V_{\min}$.
	The straight-line interpolation remains inside the
	Bloch ball and attains this minimum.
\end{proof}

This switching constraint applies when the marker
input is varied continuously while the apparatus
remains fixed. It does not apply to discontinuous
input changes or to retuning the path-dependent marker interaction.

\subsection{What a disconnected section certifies}
\label{sec:witness}

\begin{corollary}[Unital preparation excludes disconnected sections]
	\label{cor:unital}
	Assume balanced path priors ($p=1/2$), arbitrary
	fixed source coherence, arbitrary fixed
	path-dependent marker unitaries, and a common CPTP post-interaction
	marker channel. If the marker input ranges over
	the full Bloch ball and the preparation channel
	$\Phi$ is unital, every feasible fixed-visibility
	distinguishability section is a single closed
	interval. Thus a disconnected section witnesses
	nonunital marker preparation within this model.
\end{corollary}

\begin{proof}
	For a unital preparation channel, $\bv t=0$,
	so $\xi=0$. Balanced priors give $\delta=0$
	and hence $\bv d=0$, independently of $\bv h$.
	The transformation $\bv x\mapsto-\bv x$
	exchanges the two signed input disks, while
	\[
	D(-\bv x)=\norm{-A\bv x}
	=\norm{A\bv x}=D(\bv x).
	\]
	Their distinguishability intervals therefore
	coincide. A single feasible signed section is
	connected, and the degenerate cases follow
	from the full-ball or point constructions above.
\end{proof}

The converse does not hold: nonunital preparation
does not necessarily produce a gap. For example,
consider a CPTP channel with
$T=\operatorname{diag}(a,a,b)$ and
$\bv t=t\bv e_z\neq0$, with a relative rotation about the
$z$ axis and $\Psi=\mathrm{id}$.
For balanced priors,
\[
D=|a|\sin(\theta/2)
\sqrt{x_x^2+x_y^2}.
\]
Each feasible signed disk contains its center,
where $D=0$, so the two intervals overlap.

The witness also depends on the discrimination
task and the path priors. For unequal priors,
the term $\delta\bv h$ can break the sign
symmetry even when preparation is unital.
The effect of prior imbalance is considered
in Sec.~\ref{sec:certification}.
A nonunital path channel may also produce a gap
for final-path inference, as shown in
Supplemental Sec.~\ref{app:finalgap}.

Identifying a disconnected section also requires an
adequately characterized input domain or a validated
coverage bound.

\section{Geometric origin of distinguishability gaps}
\label{sec:geometry}

The preceding witness shows that nonunitality is necessary
for a gap under balanced original-path discrimination,
but not why some nonunital channels exhibit one while others do not.
To resolve this distinction, we examine the shape and principal
directions of the prepared-marker Bloch ellipsoid. The affine
map $\bv r=T\bv x+\bv t$ places its center at $\bv t$.
The matrix $S$ describes the ellipsoid, and $P_n$ projects onto
the plane transverse to the relative rotation axis:
\begin{equation}
	S=TT^{\mathsf T},\qquad
	P_n=\mathbb{I}_3-\bv n\bv n^{\mathsf T}.
	\label{eq:shape}
\end{equation}
The eigenvalues of $S$ give the squared semiaxis lengths
of the prepared-marker Bloch ellipsoid; an orthonormal
eigenbasis specifies its principal directions. This also
covers degenerate ellipsoids with vanishing semiaxes.
\subsection{Principal axes and a gap-existence criterion}

\begin{proposition}[Rotation about a principal axis excludes gaps]
	\label{prop:principalaxis}
	For balanced path priors and original-path
	discrimination, suppose
	$S\bv n=G^2\bv n$ with $G>0$.
	Every feasible fixed-visibility section is
	connected, regardless of the preparation
	displacement, relative rotation angle, source
	coherence, or fixed common CPTP post-interaction
	marker channel.
\end{proposition}

For a principal relative rotation axis, the two signed input disks have
identical transverse image centers and linear maps. Their radii
may differ, but the resulting distinguishability intervals are
nested, so their union is connected. The proof is given in
Supplemental Sec.~\ref{supp:principalproof}.

A necessary and sufficient condition for gap existence can be
established in the following restricted setting: a balanced
pure quanton with $\Psi=\mathrm{id}$,
$U_1=\mathbb{I}_2$, and
$U_0=-i\,\bv n\cdot\bv\sigma$, as in
Eq.~\eqref{eq:geometryVD}. In this setting,
$\bv r_\perp=P_n\bv r$.

\begin{theorem}[Gap existence for perpendicular rotation axes]
	\label{thm:geometry}
	Let $\bv t\ne0$ and restrict the relative rotation axes
	to $\bv n\cdot\bv t=0$.
	In the preceding setting, some such axis
	produces a disconnected fixed-visibility
	section if and only if the displacement $\bv t$ is not
	an eigenvector of the ellipsoid shape matrix $S$ (i.e.,
	it does not lie along a principal direction), equivalently
	\begin{equation}
		S\bv t\notin\operatorname{span}\{\bv t\}.
		\label{eq:geometrycriterion}
	\end{equation}
	When this condition holds, a suitable axis is
	\begin{equation}
		\bv w=
		\left(\mathbb{I}_3-\frac{\bv t\bv t^{\mathsf T}}
		{\norm{\bv t}^2}\right)S\bv t,
		\qquad
		\bv n=\frac{\bv w}{\norm{\bv w}}.
		\label{eq:constructaxis}
	\end{equation}
	For this axis, the fixed-visibility sections
	are disconnected throughout an open interval
	immediately below
	$G=\sqrt{\bv n^{\mathsf T}S\bv n}$.
\end{theorem}

\begin{proof}
	Since $\bv n\cdot\bv t=0$, the axial
	projection ranges over $[-G,G]$.
	At $V=G>0$, the signed input disks
	collapse to the points
	\[
	\bv x_\pm=
	\pm\frac{T^{\mathsf T}\bv n}{G}.
	\]
	The corresponding output Bloch vectors
	and distinguishabilities are
	\begin{equation}
		\begin{gathered}
			\bv r_\pm=\bv t\pm\frac{S\bv n}{G},\\
			D_\pm=\norm{\bv t\pm\bv q},
			\qquad
			\bv q=\frac{P_nS\bv n}{G}.
		\end{gathered}
		\label{eq:geometryendpoints}
	\end{equation}
	Because $\bv n\cdot\bv t=0$,
	\begin{equation}
		D_+^2-D_-^2
		=\frac{4\bv t^{\mathsf T}S\bv n}{G}.
		\label{eq:geometrysplit}
	\end{equation}
	
	If Eq.~\eqref{eq:geometrycriterion} holds,
	the vector $\bv w$ in
	Eq.~\eqref{eq:constructaxis} is nonzero.
	The constructed axis satisfies
	$\bv n\cdot\bv t=0$ and
	\[
	\bv t^{\mathsf T}S\bv n=\norm{\bv w}>0.
	\]
	Moreover, $G>0$: otherwise, positivity
	of $S$ would imply $S\bv n=0$,
	contradicting the preceding inequality.
	Thus $D_+\ne D_-$.
	
	As $V$ approaches $G$ from below,
	the signed-disk centers approach their
	respective endpoint inputs while their
	radii vanish continuously. Their
	distinguishability intervals therefore
	remain separated for $V<G$ sufficiently
	close to $G$.
	
	Conversely, suppose
	$S\bv t=\lambda\bv t$.
	Set
	\[
	\bv e=\frac{\bv t}{\norm{\bv t}},
	\qquad
	R_\pi=2\bv e\bv e^{\mathsf T}-\mathbb{I}_3.
	\]
	The matrix $R_\pi$ is a rotation by
	$\pi$ about the displacement direction.
	Since $S$ is symmetric and $\bv t$
	is its eigenvector,
	\[
	R_\pi S R_\pi^{\mathsf T}=S,
	\qquad
	R_\pi\bv t=\bv t.
	\]
	The rotation therefore preserves the
	output Bloch ellipsoid, including when
	it is degenerate.
	
	For any $\bv n\perp\bv t$,
	$R_\pi\bv n=-\bv n$, so the rotation
	exchanges the positive and negative
	axial sections. Since $R_\pi$ commutes
	with $P_n$,
	\[
	\norm{P_nR_\pi\bv r}
	=\norm{R_\pi P_n\bv r}
	=\norm{P_n\bv r}.
	\]
	Thus the two sections have identical
	distinguishability intervals and cannot
	form a gap. When $G=0$, visibility is
	independent of the marker input and
	the section is connected.
\end{proof}

Theorem~\ref{thm:geometry} identifies the additional geometric
condition in this balanced pure-source model: a suitable
perpendicular axis exists precisely when the displacement is
not a principal direction of the ellipsoid. The result does
not assert that every perpendicular axis gives a gap.
The channels in Sec.~\ref{sec:example} illustrate how identical
displacements can accompany different attainable regions.

The endpoint argument also gives a sufficient condition
in the presence of a common post-interaction marker channel.

\begin{proposition}[Endpoint separation with post-interaction marker noise]
	\label{prop:postinteractionendpoint}
	Consider a balanced pure quanton and a relative
	$\pi$ rotation. For a fixed axis
	$\bv n\perp\bv t$ with $G>0$, let
	$L_n=BK_{-,n}$, where
	$K_{-,n}=(R_{0,n}-R_{1,n})/2$
	and $B$ is the linear part of a fixed common
	CPTP marker channel.
	
	If
	\begin{equation}
		\bv t^{\mathsf T}L_n^{\mathsf T}
		L_nS\bv n\ne0,
		\label{eq:postinteractionendpoint}
	\end{equation}
	then the fixed-visibility sections are
	disconnected throughout an open interval
	immediately below $V=G$.
\end{proposition}

The proof is given in Supplemental
Sec.~\ref{supp:noisyendpointproof}.

This condition is sufficient for the specified
axis, rather than a general necessity condition.
The map $L_n$ depends on the path-dependent marker interaction.
For $\Psi=\mathrm{id}$ and relative $\pi$ rotation,
$L_n^{\mathsf T}L_n=P_n$, recovering the endpoint
mechanism in Theorem~\ref{thm:geometry}.

The endpoint condition also gives a quantitative visibility
range over which separation is guaranteed. Let
$L=L_n$, $\bv v=L\bv t$,
$\bv c=LS\bv n/G$, and $J=LTE$, where the columns of
$E$ form an orthonormal basis of
$(T^{\mathsf T}\bv n)^\perp$.
If the two endpoint values differ by
$\Delta_e=|\norm{\bv v+\bv c}-\norm{\bv v-\bv c}|>0$,
write $s=V/G$ and $[u]_+=\max\{u,0\}$.
Here $\norm{X}_{\rm op}$ denotes the largest singular value of $X$.
Then
\begin{equation}
 \Delta D(sG)\geq
 \left[\Delta_e-2\norm{\bv c}(1-s)
 -2\norm{J}_{\rm op}\sqrt{1-s^2}\right]_+.
 \label{eq:explicitwindowbound}
\end{equation}
In particular, a gap is guaranteed for
\begin{equation}
 G\sqrt{1-\left[
 \frac{\Delta_e}{2(\norm{\bv c}+\norm{J}_{\rm op})}
 \right]^2}<V\leq G.
 \label{eq:explicitwindow}
\end{equation}
This bound is sufficient, not an exact threshold; it follows
from the shrinking disk radii and reverse triangle
inequalities (Supplemental Sec.~\ref{supp:windowproof}).

\subsection{Gap width and preparation displacement}
\label{sec:gapbound}

Beyond determining whether a gap can occur, its width can
constrain the displacement of the prepared-marker Bloch ellipsoid.
Consider balanced path priors and original-path
discrimination, with an arbitrary fixed common
CPTP post-interaction marker channel.
For $\bv n\cdot\bv t=0$ and $G>0$, define
\[
L=BK_-,
\qquad
\bv h_g=\frac{LS\bv n}{G^2}.
\]
At axial magnitude $0\leq u\leq G$,
the two signed input disks have the same
radius $R=\sqrt{1-u^2/G^2}$.
The corresponding attainable distinguishability ranges are
\begin{equation}
	I_\pm=
	\left\{
	\norm{J\bv y+L\bv t\pm u\bv h_g}:
	\norm{\bv y}\leq R
	\right\}.
	\label{eq:pairedimages}
\end{equation}

Let $\Delta D$ be the separation between
these intervals, taken as zero when they
overlap. Then
\begin{equation}
	\Delta D\leq
	2\min\{\norm{L\bv t},
	u\norm{\bv h_g}\}
	\leq2\norm{L}_{\rm op}\norm{\bv t}.
	\label{eq:gapupper}
\end{equation}

Pairing equal transverse coordinates in the two disks bounds
the difference by $2u\norm{\bv h_g}$; pairing opposite
coordinates bounds it by $2\norm{L\bv t}$.
The reverse triangle inequality then bounds the separation
of the intervals as in Eq.~\eqref{eq:gapupper}.
The derivation is in Supplemental Sec.~\ref{supp:gapboundproof}.

The singular values of the balanced
rotation difference $K_-$ are
$\sin(\theta/2)$, $\sin(\theta/2)$,
and zero. Contractivity of trace
distance under a CPTP marker channel
implies
\[
\norm{L}_{\rm op}\leq\sin(\theta/2).
\]
A positive certified gap therefore requires
$\norm{L}_{\rm op}>0$ and, for $\theta>0$,
\begin{equation}
	\norm{\bv t}
	\geq\frac{\Delta D}
	{2\norm{L}_{\rm op}}
	\geq\frac{\Delta D}
	{2\sin(\theta/2)}.
	\label{eq:gaptranslation}
\end{equation}
If $L$ is independently calibrated,
the bound involving $\norm{L}_{\rm op}$
is the stronger one.

For the static example below, the
bound based on the relative rotation angle gives
$\norm{\bv t}\geq0.0665$ at $V=0.2$,
compared with the actual displacement
$\norm{\bv t}=0.291$.

A certified gap thus provides a lower
bound on the preparation displacement
under the perpendicular-axis assumptions.
This inference does not require
independent preparation of a maximally
mixed marker, although certification
of the gap requires sufficient coverage
of the attainable section.

\section{Physical realizations of distinguishability gaps}
\label{sec:physical}

We examine the effect in two physical settings: a static
preparation channel and time-independent open-system dynamics.

\subsection{Static preparation channels with identical displacement}
\label{sec:example}

We construct a CPTP preparation channel that produces a gap,
and compare it with another channel having the same displacement
but no gap. Let $\rho_\pm$ be qubit states with Bloch vectors
$(t_0\pm v_x,0,\pm v_z)$, and consider the
measure-and-prepare channel
\begin{equation}
	\Phi_0(\rho)
	=\langle0|\rho|0\rangle\rho_+
	+\langle1|\rho|1\rangle\rho_-.
	\label{eq:measureprep}
\end{equation}
For $0<\epsilon<1$, define
\begin{equation}
	\Phi_\epsilon
	=\epsilon\,\mathrm{id}+(1-\epsilon)\Phi_0.
	\label{eq:examplechan}
\end{equation}
Both channels in this mixture are completely positive
and trace preserving. When $\rho_\pm$ are strictly
positive, the Choi matrix of $\Phi_0$ is positive
definite, and so is that of $\Phi_\epsilon$.
For an input Bloch vector $\bv x$, the measurement probabilities
are $(1\pm x_z)/2$. The output of $\Phi_0$ consequently has
Bloch vector $(t_0+v_xx_z,0,v_zx_z)$; mixing with the identity
channel produces the affine map below.

The affine Bloch representation of $\Phi_\epsilon$ is
\begin{equation}
	\begin{gathered}
		T=
		\begin{pmatrix}
			\epsilon&0&(1-\epsilon)v_x\\
			0&\epsilon&0\\
			0&0&b_0
		\end{pmatrix},
		\qquad
		b_0=\epsilon+(1-\epsilon)v_z,\\
		\bv t=((1-\epsilon)t_0,0,0).
	\end{gathered}
	\label{eq:exampleT}
\end{equation}
Choose
\begin{equation}
	\begin{gathered}
		\epsilon=0.03,\qquad t_0=0.3,\qquad
		v_x=0.2,\\
		v_z=0.4,\qquad b_0=0.418.
	\end{gathered}
	\label{eq:examplepars}
\end{equation}
The two prepared states have Bloch-vector lengths
$\sqrt{0.41}$ and $\sqrt{0.17}$, both strictly
below one. The resulting channel therefore has
a positive-definite Choi matrix. Moreover,
$\det T=0.0003762>0$, so its Bloch action is
invertible. The displacement lies along $x$,
whereas $S\bv t$ has a nonzero $z$ component.
The geometric condition in
Theorem~\ref{thm:geometry} is thus satisfied.

Take a balanced pure initial quanton state,
$U_1=\mathbb{I}_2$, $U_0=-i\,\sigma_z$, and
$\Psi=\mathrm{id}$. The visibility and
original-path distinguishability become
\begin{equation}
	\begin{gathered}
		V=b_0|x_z|,\\
		D=\sqrt{
			[\epsilon x_x+(1-\epsilon)(t_0+v_xx_z)]^2
			+\epsilon^2x_y^2}.
	\end{gathered}
	\label{eq:exampleVD}
\end{equation}
For $0\leq V\leq b_0$, define
\begin{equation}
	\begin{aligned}
		w(V)&=\epsilon\sqrt{1-(V/b_0)^2},\\
		\mu_\pm(V)&=(1-\epsilon)
		\left(t_0\pm v_x\frac{V}{b_0}\right).
	\end{aligned}
	\label{eq:centers}
\end{equation}
Each signed input disk maps to a transverse
disk of radius $w(V)$ centered at
$(\mu_\pm(V),0)$. The corresponding
distinguishability intervals are exactly
\begin{equation}
	\begin{gathered}
		I_\pm(V)=
		\left[
		\max\{0,|\mu_\pm(V)|-w(V)\},
		|\mu_\pm(V)|+w(V)
		\right],\\
		\mathcal R_V=I_-(V)\cup I_+(V).
	\end{gathered}
	\label{eq:exampleintervals}
\end{equation}

For the chosen parameters,
$\mu_-(V)\geq0.097>w(V)$ throughout
the feasible visibility range. The two
intervals are therefore separated precisely when
\begin{equation}
	V>V_{\mathrm{gap}}
	=\frac{b_0\epsilon}
	{\sqrt{\epsilon^2+(1-\epsilon)^2v_x^2}}
	\simeq0.06388.
	\label{eq:threshold}
\end{equation}
Their gap width is
\begin{equation}
	\Delta D(V)=
	2\left[
	(1-\epsilon)v_x\frac{V}{b_0}
	-\epsilon\sqrt{1-(V/b_0)^2}
	\right]_+.
	\label{eq:staticwidth}
\end{equation}
Above threshold, $\Delta D(V)$ increases
strictly with $V$ and reaches
$2(1-\epsilon)v_x=0.388$ at $V=b_0$.
At this endpoint both input disks collapse
to points. The interior example below
instead has nonzero disk radii.
A derivation is given in Supplemental
Sec.~\ref{supp:staticwidthproof}.

At $V=0.2$,
\begin{equation}
	\begin{aligned}
		\mathcal R_{0.2}\simeq{}&
		[0.1718,0.2245]\\
		&\cup[0.3575,0.4102].
	\end{aligned}
	\label{eq:numericgap}
\end{equation}
The exact expression gives
$\Delta D(0.2)=0.13295966\ldots$,
as illustrated in Fig.~\ref{fig:gap}(c).
Using Eq.~\eqref{eq:Ddef}, the corresponding optimal path-guessing
probabilities occupy approximately $[0.5859,0.6123]$ and
$[0.6787,0.7051]$. No marker input at this visibility realizes
an optimal success probability between the two intervals;
their separation is $\Delta D/2\simeq0.06648$.
The interval endpoints are attained by
pure marker inputs satisfying
\[
x_z=\pm V/b_0,\qquad
x_y=0,\qquad
x_x=\pm\sqrt{1-(V/b_0)^2}.
\]
The two branches coincide at $V=0$.
To test whether the displacement alone determines
attainability, replace $\rho_\pm$ in
Eq.~\eqref{eq:measureprep} by states with
Bloch vectors $(t_0,0,\pm v_z)$.
Keep $\epsilon$, $t_0$, $v_z$, and the
path-dependent marker interaction unchanged. The resulting
CPTP channel has
\begin{equation}
	\widetilde T=
	\operatorname{diag}(\epsilon,\epsilon,b_0),
	\qquad
	\widetilde{\bv t}=\bv t=(0.291,0,0).
	\label{eq:controlchannel}
\end{equation}
Its two signed distinguishability intervals
coincide:
\begin{equation}
	\widetilde{\mathcal R}_V=
	\left[
	(1-\epsilon)t_0-w(V),
	(1-\epsilon)t_0+w(V)
	\right].
	\label{eq:controlinterval}
\end{equation}
Here $w(V)\leq0.03<(1-\epsilon)t_0$,
so the interval endpoints remain positive.

Both channels are nonunital and have the
same displacement, giving identical outputs
for a maximally mixed marker input.
Nevertheless, only the tilted channel
has disconnected fixed-visibility sections.
Thus the displacement alone does not determine
gap formation. Removing the axial--transverse coupling
also changes the ellipsoid's principal directions and
semiaxis lengths, so this comparison does not isolate
orientation as the sole cause.
Figure~\ref{fig:control} compares the two
channels. The comparison-channel construction
is detailed in Supplemental
Sec.~\ref{sec:matchedcontrol}.

\begin{figure*}[t]
	\centering
	\includegraphics[width=0.87\linewidth]
	{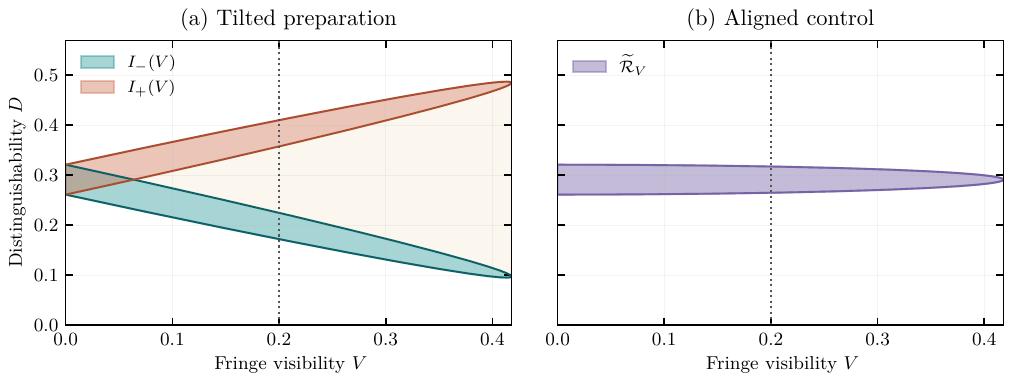}
	\caption{
		Fixed-visibility distinguishability sections
		for two preparation channels with identical
		displacement $\bv t=0.291\,\bv e_x$ and
		visibility $V=0.418|x_z|$.
		(a) The tilted-channel bands from Fig.~\ref{fig:gap}(c)
		are shown again to permit a direct comparison
		with the control channel; the inaccessible region
		is shaded amber.
		(b) Removing the axial--transverse coupling
		while retaining the same displacement
		makes the two signed intervals coincide.
		Both channels are CPTP.
		Dotted lines indicate $V=0.2$.
	}
	\label{fig:control}
\end{figure*}

\subsection{Lindblad dynamics and the gap-formation threshold}
\label{sec:markov}

The static channel above can also be obtained
at a finite time from a time-independent
Lindblad evolution. Consider the marker
dynamics
\begin{align}
	\dot r_x&=-\gamma_xr_x+
	\chi(\gamma_x-\gamma_z)r_z+t_0\gamma_x,
	\nonumber\\
	\dot r_y&=-\gamma_xr_y,\qquad
	\dot r_z=-\gamma_zr_z,
	\label{eq:markovbloch}
\end{align}
where $\gamma_x>\gamma_z>0$ and
$t_0,\chi\geq0$.
The parameter $\chi$ controls the coupling
between longitudinal and transverse Bloch
components.

These equations generate completely positive
trace-preserving dynamics of
Gorini--Kossakowski--Sudarshan--Lindblad
(GKSL) form \citep{Gorini1976,Lindblad1976}
provided
\begin{equation}
	2\gamma_x\gamma_z-\gamma_z^2
	\geq
	\chi^2(\gamma_x-\gamma_z)^2
	+t_0^2\gamma_x^2.
	\label{eq:markovphysical}
\end{equation}
The corresponding Hamiltonian, Kossakowski
matrix, and jump operators are given in
Supplemental Sec.~\ref{app:gksl}.

Writing
$a=e^{-\gamma_x\tau}$ and
$b=e^{-\gamma_z\tau}$, the $y$ and $z$ components obey
$r_y(\tau)=a x_y$ and $r_z(\tau)=b x_z$.
Substituting the latter into the equation for $r_x$
and integrating yields the cross term $\chi(b-a)x_z$
and the displacement $t_0(1-a)$. Thus the solution at time
$\tau$ is the affine map
\begin{equation}
	T(\tau)=
	\begin{pmatrix}
		a&0&\chi(b-a)\\
		0&a&0\\
		0&0&b
	\end{pmatrix},
	\qquad
	\bv t(\tau)=(t_0(1-a),0,0).
	\label{eq:markovaffine}
\end{equation}
Its linear part is invertible at every
finite time, since $\det T=a^2b>0$.

For a balanced pure quanton,
$U_0=-i\,\sigma_z$, $U_1=\mathbb{I}_2$, and
$\Psi=\mathrm{id}$, one obtains
\begin{align}
	V&=b|x_z|,\nonumber\\
	D^2&=
	[ax_x+t_0(1-a)+\chi(b-a)x_z]^2
	+a^2x_y^2.
	\label{eq:markovVD}
\end{align}

\subsubsection{Gap onset at fixed fractional visibility}

\begin{theorem}[Exact threshold for gap formation]
	\label{thm:markov}
	For the dynamics satisfying
	Eq.~\eqref{eq:markovphysical}, consider
	the source, path-dependent marker interaction, and
	post-interaction marker channel specified above.
	Fix
	$0<\zeta=V/V_{\max}(\tau)<1$,
	where $V_{\max}(\tau)=b$, and put
	$q_\zeta=\sqrt{1-\zeta^2}$.
	The complete fixed-visibility section
	is the union of
	\begin{align}
		I_\pm(\tau,\zeta)
		&=
		\left[
		\max\{0,|\mu_\pm|-aq_\zeta\},
		|\mu_\pm|+aq_\zeta
		\right],
		\nonumber\\
		\mu_\pm
		&=t_0(1-a)\pm\chi(b-a)\zeta.
		\label{eq:markovintervals}
	\end{align}
	The inaccessible-interval width is
	\begin{equation}
		\Delta D(\tau,\zeta)=
		2\max\left\{
		0,\,
		\min[t_0(1-a),\chi(b-a)\zeta]-aq_\zeta
		\right\}.
		\label{eq:markovwidth}
	\end{equation}
	If $t_0\chi>0$, the section is
	disconnected precisely when
	$\tau>\tau_c(\zeta)$, where
	\begin{equation}
		\tau_c(\zeta)=
		\max\left\{
		\frac{\ln(1+q_\zeta/t_0)}{\gamma_x},
		\frac{\ln[1+q_\zeta/(\chi\zeta)]}
		{\gamma_x-\gamma_z}
		\right\}.
		\label{eq:markovonset}
	\end{equation}
	At $\tau=\tau_c(\zeta)$, the intervals
	touch. If $t_0=0$ or $\chi=0$, no gap
	occurs in this family.
\end{theorem}

\begin{proof}
	Fixing the fractional visibility gives
	$x_z=\pm\zeta$. Each signed input disk
	has transverse radius $q_\zeta$ and
	maps to a disk of radius $aq_\zeta$
	centered at $\mu_\pm$ on the transverse
	$x$ axis. This establishes
	Eq.~\eqref{eq:markovintervals}.
	
	Let
	$d=t_0(1-a)$ and
	$h=\chi(b-a)$, consistently with Supplemental
	Sec.~\ref{app:globaloptimum}.
	Both are nonnegative. The difference
	between the absolute center distances is
	\[
	d+h\zeta-|d-h\zeta|=2\min(d,h\zeta).
	\]
	The two distinguishability intervals are
	therefore disjoint exactly when
	$\min(d,h\zeta)>aq_\zeta$.
	Their separation is
	Eq.~\eqref{eq:markovwidth}.
	
	The two strict inequalities can be written as
	\[
	t_0(e^{\gamma_x\tau}-1)>q_\zeta
	\]
	and
	\[
	\chi\zeta
	(e^{(\gamma_x-\gamma_z)\tau}-1)>q_\zeta.
	\]
	Solving them gives
	Eq.~\eqref{eq:markovonset}.
\end{proof}

\begin{figure*}[t]
	\centering
	\includegraphics[width=\linewidth]
	{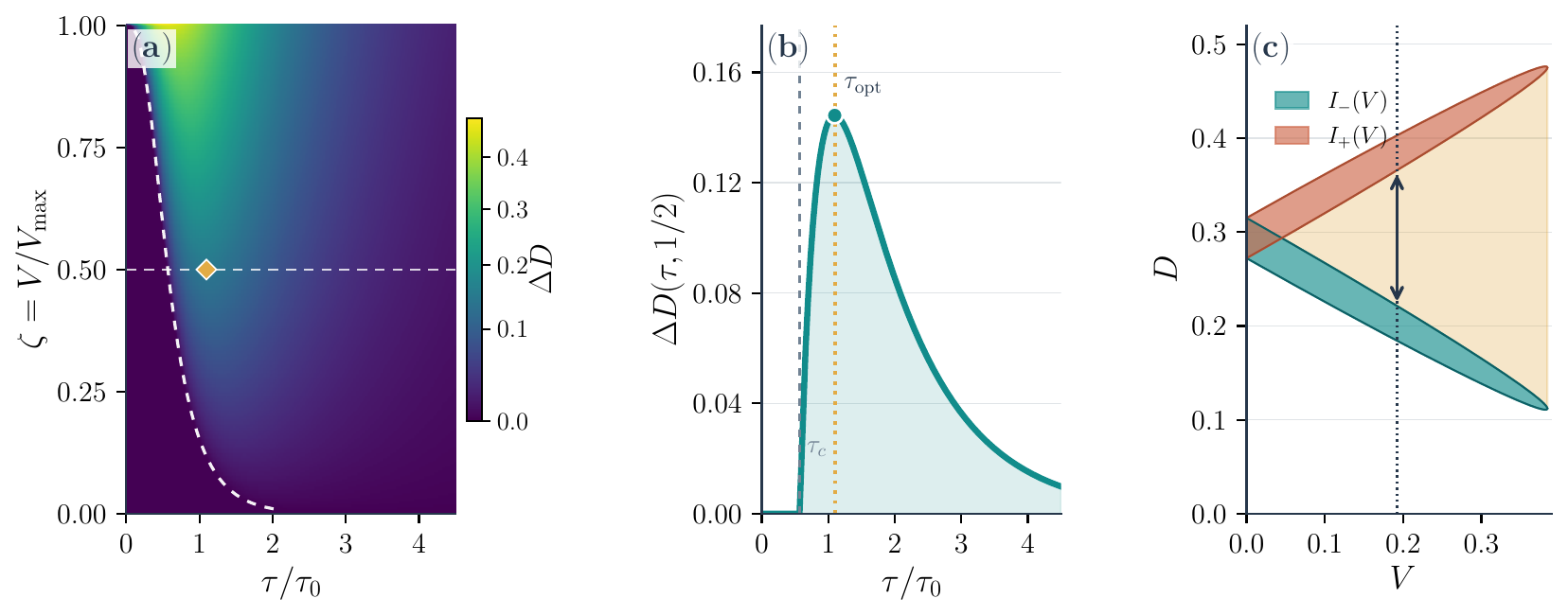}
	\caption{
		Gap formation under the time-independent
		Lindblad evolution with parameters
		in Eq.~\eqref{eq:markovparameters}.
		(a) Exact gap width $\Delta D(\tau,\zeta)$,
		where $\zeta=V/V_{\max}(\tau)$.
		The white dashed curve marks the gap-formation
		threshold, the horizontal line denotes
		$\zeta=1/2$, and the diamond indicates
		the optimal duration along that slice.
		(b) Gap width at fixed fractional visibility
		$\zeta=1/2$, with the threshold at
		$\tau_c/\tau_0\simeq0.568$ and a maximum
		$\Delta D\simeq0.1444$ at
		$\tau_{\rm opt}/\tau_0\simeq1.096$.
		(c) Complete attainable region at the
		optimal duration, with the gap at
		$V\simeq0.1922$ indicated.
		The optimization is performed at fixed
		fractional rather than fixed absolute
		visibility.
	}
	\label{fig:markov}
\end{figure*}

The critical evolution time is defined at fixed
fractional visibility. For any
$\tau>0$ and $t_0\chi>0$, a gap
occurs for sufficiently large
$\zeta<1$.

\subsubsection{Optimization at fixed fractional visibility}

To recover the static example at a
reference duration $\tau_0$, choose
\begin{align}
	\gamma_x\tau_0&=-\ln(0.03),\qquad
	\gamma_z\tau_0=-\ln(0.418),
	\nonumber\\
	t_0&=0.3,\qquad\chi=\tfrac12.
	\label{eq:markovparameters}
\end{align}
These parameters strictly satisfy
Eq.~\eqref{eq:markovphysical}.
At $\tau=\tau_0$, the affine map
in Eq.~\eqref{eq:markovaffine}
reproduces Eqs.~\eqref{eq:exampleT}
and \eqref{eq:examplepars}.

We first optimize the gap over evolution time while
following the slice $\zeta=1/2$; since
$V_{\max}(\tau)=b(\tau)$ varies, this does not
hold the absolute visibility constant.
For these parameters and $\zeta=1/2$,
\[
t_0(1-a)\geq\chi(b-a)\zeta
\]
at every evolution time. Hence
Eq.~\eqref{eq:markovwidth} reduces to
\begin{equation}
	\Delta D(\tau,1/2)=
	\max\left\{
	0,\,
	\tfrac12e^{-\gamma_z\tau}
	-(\tfrac12+\sqrt3)e^{-\gamma_x\tau}
	\right\}.
	\label{eq:markovhalfgap}
\end{equation}
The critical evolution time is
\[
\tau_c=
\frac{\ln(1+2\sqrt3)}
{\gamma_x-\gamma_z}
\simeq0.568\,\tau_0.
\]
Differentiating the positive branch
gives a unique maximizing duration,
\begin{equation}
	\tau_{\mathrm{opt}}=
	\frac{
		\ln[\gamma_x(1+2\sqrt3)/\gamma_z]
	}{\gamma_x-\gamma_z}
	\simeq1.096\,\tau_0.
	\label{eq:markovopt}
\end{equation}

The maximum gap also has a closed
analytical form. Writing
$\eta=\gamma_z/\gamma_x\in(0,1)$,
the stationary-point condition gives
\begin{equation}
	\begin{aligned}
		V_{\mathrm{opt}}
		&=\frac12\left(\frac{\eta}{1+2\sqrt3}\right)^{\eta/(1-\eta)},\\
		\Delta D_{\max}
		&=(1-\eta)V_{\mathrm{opt}}.
	\end{aligned}
	\label{eq:markovmaxexact}
\end{equation}

At $\tau_{\mathrm{opt}}$,
$V_{\mathrm{opt}}\simeq0.1922$ and
$\Delta D_{\max}\simeq0.1444$.
The corresponding section is
\begin{equation}
	\begin{aligned}
		\mathcal R_{V_{\mathrm{opt}}}
		\simeq{}&
		[0.1843,0.2214]\\
		&\cup[0.3658,0.4029].
	\end{aligned}
	\label{eq:markovnumeric}
\end{equation}
Figure~\ref{fig:markov} shows the
gap-formation boundary, the $\zeta=1/2$ slice,
and the attainable region at the
optimal duration.
Other fractional-visibility slices are treated in
Supplemental Sec.~\ref{app:generaloptimum}.
If both $\tau$ and $0\leq\zeta\leq1$ are optimized,
the largest separation for these parameters is
$0.47392250\ldots$ at $\tau/\tau_0=0.52814607\ldots$
and $\zeta=1$. At that boundary each input disk
collapses to a point: this is a maximum separation of two
attainable values, not a gap between positive-width bands
(Supplemental Sec.~\ref{app:globaloptimum}).

\subsubsection{Fixed absolute visibility and a high-visibility example}

Holding the absolute visibility $0<V<1$ fixed gives a different
time dependence. The gap is nondecreasing over the feasible evolution
times and reaches its maximum at
$\tau=\ln(1/V)/\gamma_z$, where the two signed disks
collapse to points. Supplemental Sec.~\ref{app:fixedabsolute}
gives the exact width and proof.

Gaps can also occur at appreciable
absolute visibility. Consider
\begin{equation}
	\begin{gathered}
		e^{-\gamma_x\tau_0}=0.01,\qquad
		e^{-\gamma_z\tau_0}=0.8,\\
		t_0=\tfrac14,\qquad\chi=\tfrac18.
	\end{gathered}
	\label{eq:highvisibilityparameters}
\end{equation}
This parameter choice strictly satisfies
the GKSL condition
\eqref{eq:markovphysical}.
At $\tau=\tau_0$ and $V=0.6$,
\[
\begin{aligned}
	\mathcal R_{0.6}\simeq{}&
	[0.1668,0.1801]\\
	&\cup[0.3149,0.3282].
\end{aligned}
\]
The gap width is exactly
\begin{equation}
	\Delta D(\tau_0,3/4)
	=\frac{237-8\sqrt7}{1600}
	\simeq0.13489624.
	\label{eq:highvisgap}
\end{equation}

\section{Certifying an inaccessible distinguishability interval}
\label{sec:certification}

The channel examples establish mathematical gaps over a
continuous set of marker inputs. Two clusters of experimentally
sampled outcomes would not, by themselves, establish an
inaccessible interval between them: untested inputs at the
same visibility must also be covered by calibrated bounds.

Suppose the singular values of the admissible
post-interaction marker-channel matrices $B$ lie
between $s_-$ and $s_+$. For a visibility bin
$\mathcal J$, let $U_L(\mathcal J)$ be the largest
upper edge of the lower ideal branch and
$L_H(\mathcal J)$ the smallest lower edge of
the higher branch. Allow absolute distinguishability
error $e_D$ and prior imbalance
$|\delta|\leq\delta_{\max}$. A sufficient
certification margin is
\[
\begin{aligned}
	m_{\mathrm{cert}}={}&
	s_-L_H(\mathcal J)
	-s_+U_L(\mathcal J)\\
	&-2(e_D+\delta_{\max}).
\end{aligned}
\]
When $m_{\mathrm{cert}}>0$, an open interval is
excluded throughout $\mathcal J$. To establish a
disconnected section, attainable values on both sides
must additionally be demonstrated at one common
visibility. A nonpositive margin is inconclusive.

For the static example, take
$s_-=0.8$, $s_+=1$,
$\mathcal J=[0.195,0.205]$,
$e_D=0.005$, and $\delta_{\max}=0$.
The certified exclusion contains
$(0.2321,0.2789)$, with endpoints rounded
inward; see Supplemental Fig.~\ref{fig:gapcertificate}.
Phase-resolved interference identifies the signed input section,
and path-conditioned marker measurements determine its
distinguishability.

Supplemental Secs.~\ref{sec:robust},
\ref{sec:finitecoverage}, and
\ref{sec:shotrequirements} derive the channel,
input-coverage, and measurement-resolution bounds.
A uniform conditional-state fidelity guarantee gives
another sufficient test only when the ideal and actual
fixed-visibility input sections coincide
(Supplemental Proposition~\ref{prop:fidelityrobust}).
The stated error guarantees must include marker-state
preparation, visibility calibration, and any choice of
rotation axis or evolution time based on the data.

Unlike direct nonunitality testing, which can probe
$\Phi(\mathbb{I}_2/2)$, certification of a gap concerns an entire
attainable section. The two tests therefore answer
different experimental questions.

\section{Discussion and conclusions}
\label{sec:discussion}

The usual duality inequality limits path information but does
not determine which Helstrom-optimal distinguishabilities can
be selected at a prescribed visibility. The two signed sections
make this a concrete preparation-control question. Inputs within
one section can tune distinguishability through its attainable
interval; moving to the other section can select a different
interval, but no input fills the gap between them. This
restriction applies to optimal discrimination, not to guessing
probabilities from suboptimal readout.

The result also guides the choice of preparation controls.
For a given preparation channel and path-dependent marker interaction, the endpoint
formulas identify the available distinguishability range and
inputs attaining its boundaries. If intermediate optimal path
information is needed at the same visibility, changing the
input alone may be insufficient: the preparation channel or
path-dependent marker interaction must be reconsidered. In the balanced
pure-source setting, Theorem~\ref{thm:geometry} tests whether
some rotation axis perpendicular to the preparation displacement can produce a gap. The
matched-displacement example shows why nonunitality by itself
is not a design criterion.

The fringe phase distinguishes the signed sections, and
continuous switching between them incurs the visibility cost
of Corollary~\ref{cor:switch}. Lindblad evolution supplies a
further control parameter, with different optima depending on
whether fractional or absolute visibility is held fixed.
Experimentally, phase-resolved interference is one way to
assign the branches; establishing an excluded interval also
requires calibrated coverage and error bounds, as in
Sec.~\ref{sec:certification}. Our statements concern inference
of the original path; the different final-path task after
quanton noise is discussed in the Supplemental Material.

\section*{Data and code availability}

The data and code supporting the findings of this work are
available from the corresponding author upon reasonable request. No experimental
data were collected; the finite-resolution
error estimates are illustrative.

%

\clearpage
\hypersetup{pageanchor=false}
\setcounter{page}{0}
\setcounter{section}{0}
\setcounter{subsection}{0}
\setcounter{equation}{0}
\setcounter{figure}{0}
\setcounter{table}{0}
\setcounter{theorem}{0}
\setcounter{proposition}{0}
\setcounter{corollary}{0}
\setcounter{lemma}{0}
\renewcommand{\thesection}{S\Roman{section}}
\renewcommand{\thesubsection}{\Alph{subsection}}
\renewcommand{\theequation}{S\arabic{equation}}
\renewcommand{\thefigure}{S\arabic{figure}}
\renewcommand{\thetable}{S\arabic{table}}
\renewcommand{\theproposition}{S\arabic{proposition}}
\renewcommand{\thetheorem}{S\arabic{theorem}}
\renewcommand{\theHsection}{supp.\arabic{section}}
\renewcommand{\theHsubsection}{supp.\arabic{section}.\arabic{subsection}}
\renewcommand{\theHequation}{supp.\arabic{equation}}
\renewcommand{\theHfigure}{supp.\arabic{figure}}
\renewcommand{\theHtable}{supp.\arabic{table}}
\providecommand{\theHtheorem}{}
\renewcommand{\theHtheorem}{supp.\arabic{theorem}}
\providecommand{\theHproposition}{}
\renewcommand{\theHproposition}{supp.\arabic{proposition}}
\providecommand{\theHcorollary}{}
\renewcommand{\theHcorollary}{supp.\arabic{corollary}}
\providecommand{\theHlemma}{}
\renewcommand{\theHlemma}{supp.\arabic{lemma}}
\onecolumngrid
\begin{center}
{\large\bfseries Supplemental Material: Path-Distinguishability Gaps at Fixed Visibility in Noisy Qubit Interferometers\par}
\vspace{0.7em}
Mohd Asad Siddiqui\textsuperscript{*}\\[0.3em]
{\itshape Department of Physics, Government Girls P.G. College, Sawai Madhopur, Rajasthan 322001, India}\\[0.25em]
{\footnotesize \textsuperscript{*}\href{mailto:asad@ctp-jamia.res.in}{asad@ctp-jamia.res.in}}
\end{center}
\twocolumngrid
\thispagestyle{empty}
\raggedbottom

This Supplemental Material gives the endpoint calculations and geometric proofs, a comparison of channels with identical displacement, original- and final-path discrimination formulas, the Lindblad realization, and finite-input-coverage bounds. Unless stated otherwise, distinguishability refers to the original path and visibility is evaluated before any post-interaction quanton channel; finite-resolution calculations assume calibrated error bounds rather than experimental data.
The symbol $\mathbb{I}_2$ denotes the identity on the marker qubit.

\section{Phase resolution and comparison channels}
\label{supp:phasecontrol}
\subsection{Distinguishability at fixed complex fringe amplitude}
\label{sec:phaseproof}

Let $\mathcal A(\bv x)=\alpha\Tr[\rho_\Phi(\bv x)W]$ denote the calibrated complex fringe amplitude, with the source-coherence phase removed. Since $W=\cos(\theta/2)\mathbb{I}_2-i\sin(\theta/2)\bv n\cdot\bv\sigma$, its real part is constant and its imaginary part is affine in the input Bloch vector. When $\alpha\sin(\theta/2)>0$ and $G>0$, fixing a feasible value of $\mathcal A$ imposes one real affine constraint $\bv g\cdot\bv x=\ell$. Its intersection with $\norm{\bv x}\leq1$ is a disk or a point, hence compact and connected. If $G=0$ or $\alpha\sin(\theta/2)=0$, the amplitude is independent of the marker input and the feasible input set is the whole Bloch ball. In either case the continuous Helstrom distinguishability for any fixed common post-interaction marker channel has a closed interval as its image. The argument does not require unitality or balanced path priors and also covers singular preparation maps.

For a finite-dimensional marker with a compact convex set of allowed
input density operators, the calibrated complex amplitude remains affine.
Fixing its complex value amounts to two real affine constraints, whose
intersection with the allowed states is compact and convex. The continuous
Helstrom distinguishability therefore has a closed interval as its image.
The argument need not hold for nonconvex input restrictions, including
pure marker states alone, and does not extend the qubit two-disk
characterization to arbitrary dimensions.

Phase resolution provides additional phase information for conditioning the attainable region; it does not generate new $(V,D)$ pairs. Two amplitudes $\alpha[\cos(\theta/2)\pm i\sin(\theta/2)u]$ can have the same modulus but different phases, and their phase-resolved intervals project to the two branches at fixed $V$. Determining these phases experimentally requires a calibrated phase reference.

\subsection{Nonunital preparation channels with identical displacement}
\label{sec:matchedcontrol}

In main-text Eqs.~\eqref{eq:measureprep}--\eqref{eq:examplepars}, $\Phi_\epsilon$ mixes an identity channel with a measure-and-prepare channel that sends $|0\rangle,|1\rangle$ to marker states with Bloch vectors $(t_0\pm v_x,0,\pm v_z)$. Consider instead $\widetilde\Phi_\epsilon$ constructed from Bloch vectors $(t_0,0,\pm v_z)$. For the numerical values in the main text, these have length $\sqrt{0.3^2+0.4^2}=0.5<1$, so the new measure-and-prepare channel and its convex mixture with the identity are completely positive and trace preserving (CPTP), with strictly positive output states and a full-rank Choi matrix. The linear parts satisfy
\begin{equation}
 T=\begin{pmatrix}\epsilon&0&(1-\epsilon)v_x\\0&\epsilon&0\\0&0&b_0\end{pmatrix},\qquad
 \widetilde T=\begin{pmatrix}\epsilon&0&0\\0&\epsilon&0\\0&0&b_0\end{pmatrix},
 \label{eq:ctrlT}
\end{equation}
while both have displacement $\bv t=(1-\epsilon)t_0\bv e_x$.
For the tilted example, the normalized Choi matrix is positive
definite even without numerical diagonalization: its smallest
eigenvalue obeys
$\lambda_{\min}(C_{\Phi_\epsilon})\geq
(1-\epsilon)[1-\sqrt{0.41}]/4>0$.
The singular values of $T$ and $\widetilde T$ differ, so this
is a fixed-displacement comparison rather than a controlled
change of ellipsoid orientation alone. Thus
\begin{equation}
 \Phi_\epsilon(\mathbb{I}_2/2)=\widetilde\Phi_\epsilon(\mathbb{I}_2/2)
 =\tfrac12[\mathbb{I}_2+(1-\epsilon)t_0\sigma_x].
 \label{eq:ctrlmix}
\end{equation}
For balanced paths and a relative $\pi$ rotation about $z$, their visibilities are identical: $V=b_0|x_z|$. The tilted channel has $D^2=[\epsilon x_x+(1-\epsilon)(t_0+v_xx_z)]^2+\epsilon^2x_y^2$, while the control has $\widetilde D^2=[\epsilon x_x+(1-\epsilon)t_0]^2+\epsilon^2x_y^2$. Fixing $|x_z|=V/b_0$ leaves a disk of radius $\sqrt{1-(V/b_0)^2}$. The control therefore produces the single interval in Eq.~\eqref{eq:controlinterval}; for the tilted channel its branch centers are displaced by opposite $\pm(1-\epsilon)v_xV/b_0$, generating the gap. In particular, $\widetilde S\bv t\parallel\bv t$ while $S\bv t\not\parallel\bv t$.

\subsection{Exact gap width for the static channel}
\label{supp:staticwidthproof}
For the parameters of main-text Eq.~\eqref{eq:examplepars}, the disk centers obey $\mu_+(V)\geq\mu_-(V)>w(V)$ for $0\leq V\leq b_0$. The upper edge of the lower branch and the lower edge of the higher branch are therefore $U_L=\mu_-+w$ and $L_H=\mu_+-w$. Their difference gives Eq.~\eqref{eq:staticwidth}; when the difference is nonpositive the two intervals overlap. For $V_{\rm gap}<V<b_0$, differentiation gives
\begin{equation}
 \frac{d\Delta D}{dV}=2\left[\frac{(1-\epsilon)v_x}{b_0}
 +\frac{\epsilon V}{b_0^2\sqrt{1-(V/b_0)^2}}\right]>0.
 \label{eq:staticderivative}
\end{equation}
Thus the maximum occurs at $V=b_0$, where both disks collapse to points and $\Delta D=2(1-\epsilon)v_x$. The exact gap endpoints for any $V>V_{\rm gap}$ are $\mu_-(V)+w(V)$ and $\mu_+(V)-w(V)$, respectively. At $V=0.2$, these are approximately $0.22452017$ and $0.35747983$. The inward-rounded interval $(0.2246,0.3574)$ is therefore a conservative example of exclusion, not a separate attainable-section calculation.

\subsection{Comparison with direct nonunitality testing}
\label{sec:resourcecomparison}

Direct detection of nonunitality for a trusted qubit preparation channel needs only the maximally mixed input and the three Pauli expectation values of its output: $\bv t=\Tr[\Phi(\mathbb{I}_2/2)\bv\sigma]$. A nonzero component, with a certified error interval excluding zero, already rejects unitality. In contrast, a gap demonstration characterizes visibility and path-conditioned marker discrimination \emph{over a continuous input domain}; with finite control and data it requires coverage assumptions and simultaneous confidence bounds from Sec.~\ref{sec:finitecoverage}. The two tasks therefore have different measurement requirements; comparing their sample costs requires a specified experimental resource model.

The direct displacement measurement does not determine the ellipsoid's shape: Eq.~\eqref{eq:ctrlmix} shows two channels with identical direct displacement readouts and different attainable fixed-$V$ sections. Full channel tomography, which also determines $T$, can of course recover that difference. Thus the predicted gap is a \emph{task-dependent geometric signature}, not a generally privileged test of nonunitality. In an input-selection task at prescribed visibility, it predicts which \emph{optimal} path-discrimination success probabilities are attainable; intermediate probabilities obtained using intentionally suboptimal measurements do not fill a gap in the Helstrom-optimum region.

\section{Endpoint optimization and geometric bounds}\label{supp:endpoints}
\subsection{Quantitative endpoint-separation window}
\label{supp:windowproof}
For a balanced pure quanton and relative $\pi$ marker rotation,
fix $\bv n\perp\bv t$ with $G=\norm{T^{\mathsf T}\bv n}>0$.
Using the definitions before main-text
Eq.~\eqref{eq:explicitwindowbound}, the two signed disks
at $V=sG$ have the following distinguishability ranges:
\begin{equation}
 \left\{\norm{\bv v\pm s\bv c+J\bv y}:
 \norm{\bv y}\leq\sqrt{1-s^2}\right\},
 \qquad 0\leq s\leq1.
 \label{eq:windowdisks}
\end{equation}
Here $\norm{J}_{\rm op}$ is the largest singular value of $J$,
and $[x]_+=\max\{x,0\}$ denotes the positive part.
Their radius-dependent deviations from the respective
center norms are at most
$\norm{J}_{\rm op}\sqrt{1-s^2}$, and each center norm
changes from its value at $s=1$ by at most
$\norm{\bv c}(1-s)$. Thus the distance between the
two distinguishability intervals, taken as zero when they overlap,
is at least
\begin{equation}
 \left[\Delta_e-2\norm{\bv c}(1-s)
 -2\norm{J}_{\rm op}\sqrt{1-s^2}\right]_+,
 \label{eq:windowproofbound}
\end{equation}
where
$\Delta_e=|\norm{\bv v+\bv c}-\norm{\bv v-\bv c}|$.
Since $1-s\leq\sqrt{1-s^2}$ for $s\in[0,1]$, the
lower bound is positive whenever
$\sqrt{1-s^2}<\Delta_e/[2(\norm{\bv c}+
\norm{J}_{\rm op})]$. The right-hand ratio is in $(0,1]$
when $\Delta_e>0$, by the reverse triangle inequality.
This proves main-text Eq.~\eqref{eq:explicitwindow}.
The result is a conservative sufficient condition and also
applies with a fixed post-interaction marker channel through $L$.

\subsection{Quadratic endpoints}\label{app:trust}
\label{sec:trust}
The interval endpoints follow by minimizing and maximizing a quadratic function over one input disk, a standard trust-region problem. For one slice let $\mathsf{Q}=J^{\mathsf T}J$, $\bv q=J^{\mathsf T}\bv e$, and $k=\norm{\bv e}^2$, so $f(\bv y)=\bv y^{\mathsf T}\mathsf{Q}\bv y+2\bv q\cdot\bv y+k$. Here $f^-$ and $f^+$ denote the minimum and maximum endpoints,
whereas a superscript $+$ on a matrix denotes its Moore--Penrose
pseudoinverse. The Lagrange multipliers enforce the disk constraint:
$\mu=0$ is permitted for an interior minimum, whereas an
active boundary constraint requires $\norm{\bv y}=R$.
A nonconstant convex quadratic reaches its maximum on the
boundary. For $R>0$, the standard trust-region dual formulas
give explicit endpoint certificates \citep{SUPPMore1983}:
\begin{align}
 f^-&=k+\max_{\mu\geq0}
       \{-\mu R^2-\bv q^{\mathsf T}(\mathsf{Q}+\mu\mathbb{I}_2)^+\bv q\},
 \label{eq:dualmin}\\
 f^+&=k+\min_{\substack{\nu\geq\lambda_{\max}(\mathsf{Q})\\
                   \bv q\cdot\bv v=0\ \mathrm{if}\ \mathsf{Q}\bv v=\nu\bv v}}
       \{\nu R^2+\bv q^{\mathsf T}(\nu\mathbb{I}_2-\mathsf{Q})^+\bv q\}.
 \label{eq:dualmax}
\end{align}
The condition in Eq.~\eqref{eq:dualmax} requires $\bv q$ to be
orthogonal to every zero-eigenvalue eigenvector of $\nu\mathbb{I}_2-\mathsf Q$
(equivalently, every eigenvector of $\mathsf Q$ with eigenvalue $\nu$).
It restricts only the endpoint $\nu=\lambda_{\max}(\mathsf Q)$;
all larger $\nu$ are allowed. The minimum formula is valid for every
$\mu\geq0$, as proved below. For $R=0$, both endpoints equal $k$.

For the minimum, if $\norm{\mathsf{Q}^+\bv q}\leq R$, then $\bv y^-=-\mathsf{Q}^+\bv q$ attains the minimum. Otherwise there is a unique $\mu>0$ satisfying
\begin{equation}
 \norm{(\mathsf{Q}+\mu\mathbb{I}_2)^{-1}\bv q}=R,
 \qquad \bv y^-=-(\mathsf{Q}+\mu\mathbb{I}_2)^{-1}\bv q.
 \label{eq:minroot}
\end{equation}
For a regular maximum there is a unique $\nu>\lambda_{\max}(\mathsf{Q})$ with
\begin{equation}
 \norm{(\nu\mathbb{I}_2-\mathsf{Q})^{-1}\bv q}=R,
 \qquad \bv y^+=(\nu\mathbb{I}_2-\mathsf{Q})^{-1}\bv q.
 \label{eq:maxroot}
\end{equation}
An exceptional maximum, called the trust-region hard case, occurs at $\nu=\lambda_{\max}(\mathsf{Q})$: if $\bv q$ is orthogonal to all eigenvectors with the largest eigenvalue and $\norm{\bv y_0}\leq R$ for $\bv y_0=(\nu\mathbb{I}_2-\mathsf{Q})^+\bv q$, choose
\begin{equation}
 \bv y^+=\bv y_0+\bv v,\quad
 \mathsf Q\bv v=\nu\bv v,\quad
 \norm{\bv v}^2=R^2-\norm{\bv y_0}^2.
 \label{eq:hard}
\end{equation}
Equations~\eqref{eq:dualmin}--\eqref{eq:hard} include singular matrices and the exceptional maximum. They give the endpoint values in Theorem~\ref{thm:region} and input states attaining them. The following argument proves the endpoint bounds and their attainability.

\subsubsection{Proof of the endpoint bounds}

For Hermitian matrices, $M\succeq0$ means that $M$ is positive
semidefinite, and $M\preceq N$ means that $N-M\succeq0$.
For the minimum and $\mu\geq0$, complete the square in
\begin{equation}
 f(\bv y)+\mu(\norm{\bv y}^2-R^2).
\end{equation}
If $\mathsf Q\bv v=0$, then $\norm{J\bv v}^2=\bv v^{\mathsf T}\mathsf Q\bv v=0$. Hence $J\bv v=0$ and $\bv v^{\mathsf T}\bv q=(J\bv v)^{\mathsf T}\bv e=0$. Thus $\bv q$ is orthogonal to every eigenvector of $\mathsf Q$ with zero eigenvalue. For $\mu>0$, $\mathsf{Q}+\mu\mathbb{I}_2$ is invertible. Hence the infimum over $\bv y$ equals $k-\mu R^2-\bv q^{\mathsf T}(\mathsf{Q}+\mu\mathbb{I}_2)^+\bv q$ for every $\mu\geq0$. It bounds $f$ below on the disk. If $\norm{\mathsf{Q}^+\bv q}\leq R$, equality holds at $\mu=0$; otherwise Eq.~\eqref{eq:minroot} attains it. The positive root is unique, since $\norm{(\mathsf{Q}+\mu\mathbb{I}_2)^{-1}\bv q}$ decreases strictly from above $R$ to zero.

For the maximum, $\nu\mathbb{I}_2-\mathsf{Q}\succeq0$ implies
\begin{align}
 f(\bv y)&=k+\nu\norm{\bv y}^2
       -\bv y^{\mathsf T}(\nu\mathbb{I}_2-\mathsf{Q})\bv y+2\bv q\cdot\bv y\nonumber\\
 &\leq k+\nu R^2+\bv q^{\mathsf T}(\nu\mathbb{I}_2-\mathsf{Q})^+\bv q
 \label{eq:maxcertificate}
\end{align}
for $\nu>\lambda_{\max}(\mathsf{Q})$, or at equality when $\bv q$ is orthogonal to all eigenvectors with the largest eigenvalue. The regular root of Eq.~\eqref{eq:maxroot} attains equality. If the norm at the singular endpoint can be filled by a vector along eigenvectors with the largest eigenvalue, Eq.~\eqref{eq:hard} attains equality instead. If that norm is already larger than $R$, or $\bv q$ has a component along eigenvectors with the largest eigenvalue, the regular root exists and is unique. The regular and hard cases together cover $\mathsf{Q}=0$, $\bv q=0$, and singular $\mathsf{Q}$. For a full-ball degeneracy in main-text Sec.~\ref{sec:region}, the same argument holds in dimension three.

\subsection{Principal-axis exclusion}
\label{supp:principalproof}
We prove main-text Proposition~\ref{prop:principalaxis}.
\begin{proof}
	For balanced priors, the rotation difference
	$K_-=(R_0-R_1)/2$ satisfies $K_-\bv n=0$.
	Consequently,
	\[
	A\bv g=BK_-S\bv n
	=G^2BK_-\bv n=0.
	\]
	The affine centers of the signed-disk images
	in Theorem~\ref{thm:region} therefore coincide:
	\[
	\bv e_\sigma=A\bv x_\sigma+\bv d
	=BK_-\bv t.
	\]
	Their linear map $J$ is also the same.
	Only the disk radii may differ, so the two
	disk domains are nested. Their continuous
	distinguishability ranges are consequently
	nested intervals, whose union is connected.
	The degenerate cases follow from the full-ball
	and single-point constructions.
\end{proof}

\subsection{Endpoint separation after marker noise}
\label{supp:noisyendpointproof}
We prove main-text Proposition~\ref{prop:postinteractionendpoint}.
\begin{proof}
	At $V=G$ the two input disks collapse, and
	balanced priors eliminate the marker-channel
	translation $\bv h$. The endpoint values obey
	\[
	D_+^2-D_-^2=
	\frac{4\bv t^{\mathsf T}
		L_n^{\mathsf T}L_nS\bv n}{G}.
	\]
	Under the stated condition they differ.
	The images of the two disks shrink
	continuously to these endpoints, leaving
	a gap for $V<G$ sufficiently close to $G$.
\end{proof}

\subsection{Gap-width bound}
\label{supp:gapboundproof}
Use the balanced-prior, perpendicular-axis assumptions and
the paired interval images in main-text Eq.~\eqref{eq:pairedimages}.
To prove the bound, compare points with the
same transverse coordinate $\bv y$ in the
two signed disks. The reverse triangle
inequality gives
\[
\begin{aligned}
	&\left|
	\norm{J\bv y+L\bv t+u\bv h_g}
	-
	\norm{J\bv y+L\bv t-u\bv h_g}
	\right|\\
	&\hspace{2cm}\leq2u\norm{\bv h_g}.
\end{aligned}
\]
Alternatively, comparing $\bv y$ in one
disk with $-\bv y$ in the other gives
\[
\begin{aligned}
	&\left|
	\norm{J\bv y+L\bv t+u\bv h_g}
	-
	\norm{-J\bv y+L\bv t-u\bv h_g}
	\right|\\
	&\hspace{2cm}\leq2\norm{L\bv t}.
\end{aligned}
\]
Each comparison bounds the difference
between corresponding interval minima
and maxima, and hence any separation
between the intervals. This proves
Eq.~\eqref{eq:gapupper}.

\section{Robustness and finite-coverage certification}\label{supp:certification}
\subsection{Post-interaction channel uncertainty and finite resolution}
\label{sec:robust}
The general construction allows arbitrary $B$ and $\bv h$. A simpler sufficient certificate is useful when only bounds on $B$ are known. The certification margin below may be negative, unlike the nonnegative certified exclusion width.

Let $D_{\mathrm{bal}}$ use balanced priors with the same marker states and source coherence. The trace-norm triangle inequality gives
\begin{equation}
 \begin{aligned}
 |D-D_{\mathrm{bal}}|
 &\leq\left\lVert\frac{\delta}{2}(\rho_0+\rho_1)\right\rVert_1\\
 &=|\delta|.
 \end{aligned}
 \label{eq:priorperturbation}
\end{equation}
At fixed $V$ and $\alpha$, the input disks are unchanged, so each distinguishability endpoint moves by at most $|\delta|$. For unital preparation the balanced intervals coincide by Corollary~\ref{cor:unital}; a gap created solely by $|\delta|\leq\delta_{\max}$ has width at most $2\delta_{\max}$. This statement concerns the path before the interaction. Uncertainty in $\alpha$ must additionally enter the visibility slices or bin.

\begin{proposition}[Post-interaction channel and finite-resolution certificate]
\label{prop:robust}
At a fixed visibility and source coherence, let the balanced-reference intervals before the post-interaction marker channel be $[L_L,U_L]$ and $[L_H,U_H]$, with $U_L<L_H$. Allow actual priors satisfying $|\delta|\leq\delta_{\max}$. Suppose every admissible post-interaction marker channel satisfies
\begin{equation}
 s_-\norm{\bv v}\leq\norm{B\bv v}\leq s_+\norm{\bv v}
 \quad\text{for all }\bv v,\qquad 0\leq s_-\leq s_+.
 \label{eq:singular}
\end{equation}
For an admissible family $\mathcal B$, one may take $s_-=\inf_{B\in\mathcal B}\sigma_{\min}(B)$ and $s_+=\sup_{B\in\mathcal B}\sigma_{\max}(B)$; any certified weaker uniform bounds are also valid. For exactly balanced priors, $s_-L_H-s_+U_L>0$ excludes $(s_+U_L,s_-L_H)$ after the channel, regardless of its translation or orientation. For a visibility bin $\mathcal J$, define
\begin{equation}
 U_L(\mathcal J)=\sup_{V\in\mathcal J}U_L(V),\qquad
 L_H(\mathcal J)=\inf_{V\in\mathcal J}L_H(V),
 \label{eq:bin}
\end{equation}
assuming the two branch labels remain valid throughout $\mathcal J$. If the measured distinguishability has absolute error at most $e_D$, then the interval
\begin{equation}
 \begin{split}
 (&s_+U_L(\mathcal J)+e_D+\delta_{\max},\\
  &s_-L_H(\mathcal J)-e_D-\delta_{\max})
 \end{split}
 \label{eq:measuredgap}
\end{equation}
is inaccessible to measured distinguishabilities whenever the certification margin
\begin{equation}
 m_{\mathrm{cert}}=s_-L_H(\mathcal J)-s_+U_L(\mathcal J)-2(e_D+\delta_{\max})
 \label{eq:certgap}
\end{equation}
is positive. The resulting certified exclusion width is
$\Delta D_{\mathrm{cert}}=\max\{0,m_{\mathrm{cert}}\}$; when the margin is nonpositive, the sufficient test is inconclusive.
\end{proposition}

\begin{proof}
For balanced priors the pre-channel difference vector is $\bv q=(R_0-R_1)\bv r/2$, and $D_{\mathrm{bal}}=\norm{B\bv q}$ after the marker channel. Thus the reference branches lie inside $[s_-L_L,s_+U_L]$ and $[s_-L_H,s_+U_H]$. Equation~\eqref{eq:priorperturbation} expands each by at most $\delta_{\max}$. Taking adverse extrema over $\mathcal J$ and allowing measurement error $e_D$ proves the statement without assumptions on branch directions.
\end{proof}

The certificate is sufficient, not necessary: a particular post-interaction marker channel may preserve a gap even when the margin is nonpositive. It also specifies only an inaccessible interval, not the exact noisy endpoints. Complete endpoints for a known post-interaction marker channel are obtained from Theorem~\ref{thm:region}. Trace-distance contractivity permits choosing $s_+=1$ for any family of physical qubit post-interaction marker channels (or a tighter certified uniform upper bound). If $s_-=0$, this certificate cannot be positive.

\begin{figure}[t]
 \centering
 \includegraphics[width=\columnwidth]{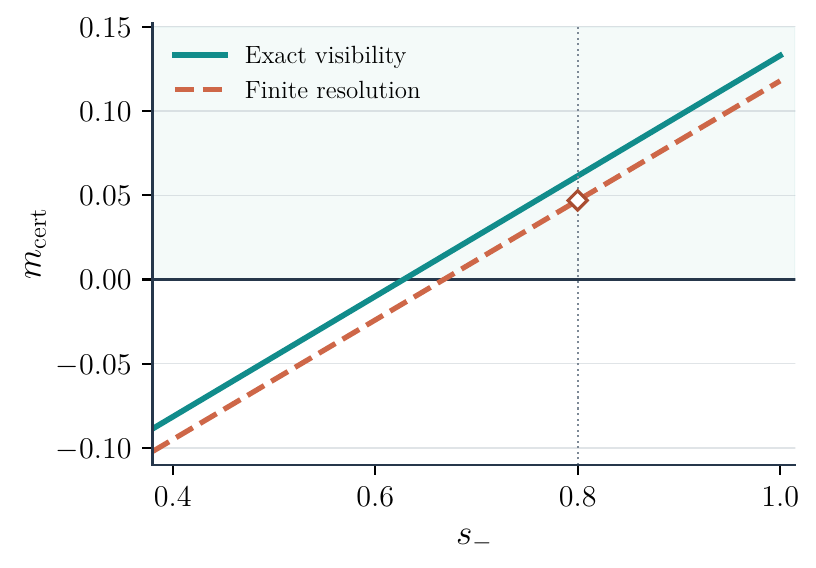}
 \caption{Certification margin for the static channel, with $s_+=1$. The solid curve uses exact visibility $V=0.2$ and zero distinguishability error; the dashed curve allows $V\in[0.195,0.205]$ and $e_D=0.005$. The open diamond marks $s_-=0.8$. A positive margin gives the sufficient exclusion in Proposition~\ref{prop:robust}; a nonpositive margin is inconclusive.}
 \label{fig:gapcertificate}
\end{figure}

For the balanced static channel and $s_+=1$, the sufficient channel-contraction threshold has the closed form
\begin{equation}
 s_-^{\mathrm{crit}}(V)=\frac{\mu_-(V)+w(V)}{\mu_+(V)-w(V)}
 \quad(V>V_{\mathrm{gap}}),
 \label{eq:staticrobustthreshold}
\end{equation}
with certification for $s_->s_-^{\mathrm{crit}}$. At $V=0.2$, the threshold is $0.6280638757\ldots$. For a visibility bin $\mathcal J=[V_1,V_2]\subset(V_{\mathrm{gap}},b_0]$, the function $U_L(V)=\mu_-(V)+w(V)$ decreases with $V$, while $L_H(V)=\mu_+(V)-w(V)$ increases. Therefore the adverse extrema in Eq.~\eqref{eq:bin} are both attained at $V_1$, and Eq.~\eqref{eq:certgap} becomes
\begin{equation}
 \begin{aligned}
 m_{\mathrm{cert}}={}&s_-[\mu_+(V_1)-w(V_1)]\\
 &-s_+[\mu_-(V_1)+w(V_1)]
 -2(e_D+\delta_{\max}).
 \end{aligned}
 \label{eq:staticcertanalytic}
\end{equation}
For $s_-=0.8$, $s_+=1$, $\mathcal J=[0.195,0.205]$, $e_D=0.005$, and $\delta_{\max}=0$, this gives $m_{\rm cert}=0.0469403659\ldots$. The exact analytically certified inaccessible interval is $(0.2320331300\ldots,0.2789734960\ldots)$, of which $(0.2321,0.2789)$ is a strictly contained four-decimal interval. At the Markovian optimum, the same $s_-$ and $e_V=e_D=0.005$ give the certified interval $(0.2288981126\ldots,0.2855998697\ldots)$, containing $(0.2290,0.2855)$; this second evaluation uses the Lindblad slice, not Eq.~\eqref{eq:staticcertanalytic}. Supplemental Fig.~\ref{fig:gapcertificate} and these intervals use $\delta_{\max}=0$; an allowed prior imbalance shrinks each end by $\delta_{\max}$. Visibility and source-coherence uncertainties must enter the bin or fixed-visibility input slices. No alignment of the post-interaction marker channel is required.

The separation is also stable under parameter perturbations that preserve both feasible signed slices. Write $\mathbb B_R=\{\bv y:\norm{\bv y}\leq R\}$ for a disk of radius $R$. If its affine image $\bv e+J\mathbb B_R$ changes to $\bv e'+J'\mathbb B_{R'}$, its Hausdorff distance is at most
\begin{equation}
 \epsilon_A=\norm{\bv e-\bv e'}+R\norm{J-J'}_{\rm op}
 +\norm{J'}_{\rm op}|R-R'|.
 \label{eq:diskstability}
\end{equation}
Pairing $R\bv z$ with $R'\bv z$ for $\norm{\bv z}\leq1$ proves the bound in both directions. Since the norm is 1-Lipschitz, each endpoint changes by at most $\epsilon_A$. A gap $\Delta D$ therefore retains width at least $\max\{0,\Delta D-\epsilon_{A,-}-\epsilon_{A,+}\}$, provided branch feasibility and ordering are maintained. Near collapsed disks, radius changes can scale as the square root of parameter error; Eq.~\eqref{eq:diskstability} should be evaluated rather than replaced by an assumed uniform linear tolerance.

We use the squared fidelity
$F_s(\rho,\sigma)=F(\rho,\sigma)^2$, where
$F(\rho,\sigma)=\|\sqrt{\rho}\sqrt{\sigma}\|_1$
is the root fidelity.
\begin{proposition}[Uniform fidelity-error exclusion]
\label{prop:fidelityrobust}
At every marker input on a calibrated fixed-visibility section,
let $\rho_j$ be the ideal normalized conditional marker state
and $\widetilde\rho_j$ its actual counterpart, for $j=0,1$.
Suppose the squared Uhlmann fidelities obey
$F_s(\rho_j,\widetilde\rho_j)\geq1-\varepsilon_j$
uniformly over the section. At the same path priors,
\[
 |\widetilde D-D|\leq
 2p\sqrt{\varepsilon_0}+2(1-p)\sqrt{\varepsilon_1}.
\]
If both errors are at most $\varepsilon$, and the ideal section
has two separated branches with edges $U_L<L_H$, every
actual distinguishability lies outside
$(U_L+2\sqrt\varepsilon,L_H-2\sqrt\varepsilon)$.
This interval is nonempty when $L_H-U_L>4\sqrt\varepsilon$.
The claim requires the same allowed marker inputs and priors
for the ideal and actual configurations, coincident fixed-visibility
input sections, and both branches to remain feasible.
If their visibility functions differ, the statement does not apply
directly; visibility uncertainty must be treated using a calibrated band.
\end{proposition}
\begin{proof}
The trace-distance--fidelity inequality\citep{SUPPFuchsVanDeGraaf1999} gives
$\|\widetilde\rho_j-\rho_j\|_1\leq2\sqrt{\varepsilon_j}$.
The reverse triangle inequality applied to the two
Helstrom operators proves the stated bound. Each ideal
branch is then enlarged by at most $2\sqrt\varepsilon$ on
each side, proving the excluded interval.
\end{proof}

Fidelity constraints are also used in prepare-and-measure
certification~\citep{SUPPdeOliveiraJunior2026}, but the present bound
assumes trusted, uniformly calibrated conditional states; it is
not a semi-device-independent certification statement. For the
static example at $V=0.2$, where $L_H-U_L\simeq0.13296$,
the sufficient criterion is $\varepsilon<1.105\times10^{-3}$.
This fidelity test controls distinguishability errors on a known
input section; it does not by itself establish that the section has
been covered experimentally. The channel-contraction and finite-input
coverage bounds address those separate calibration requirements.

\subsection{Finite input coverage}
\label{sec:finitecoverage}
A finite scan can exclude untested inputs when their outputs are bounded by a calibrated covering of the input domain. Let $\{\bv x_i\}$ be an $h_{\mathrm{net}}$-net of the allowed Bloch ball: every allowed input lies within Euclidean distance $h_{\mathrm{net}}$ of at least one $\bv x_i$. Suppose the estimated pairs $(\widehat V_i,\widehat D_i)$ have simultaneous errors bounded by $e_V$ and $e_D$. For the visibility defined before post-interaction quanton noise, the Lipschitz constants are
\begin{equation}
 L_V=\alpha\sin(\theta/2)\norm{T^{\mathsf T}\bv n}\leq1,
 \qquad L_D=\norm{BK_- T}_{\mathrm{op}}\leq1
 \label{eq:lipschitz}
\end{equation}
These constants satisfy $|V(\bv x)-V(\bv y)|\leq L_V\norm{\bv x-\bv y}$ and $|D(\bv x)-D(\bv y)|\leq L_D\norm{\bv x-\bv y}$. The reverse triangle inequality gives these estimates; taking the maximum with $P$ is also 1-Lipschitz. Trace-distance contractivity bounds $\norm{T}_{\rm op}$ and $\norm{B}_{\rm op}$ by one, while $\norm{K_-}_{\rm op}\leq1$. At target visibility $v$, form
\begin{equation}
 \begin{split}
 \mathcal I_v&=\{i:|\widehat V_i-v|\leq L_Vh_{\mathrm{net}}+e_V\},\\
 \mathcal O_v&=\bigcup_{i\in\mathcal I_v}
 [\widehat D_i-L_Dh_{\mathrm{net}}-e_D,\widehat D_i+L_Dh_{\mathrm{net}}+e_D].
 \end{split}
 \label{eq:finiteouter}
\end{equation}
Every attainable $D$ at exactly $V=v$ lies in $\mathcal O_v$. Any open interval disjoint from this outer set is excluded under the calibrated coverage and error assumptions below. Full three-dimensional coverage can be costly (the number of net points typically scales as $h_{\mathrm{net}}^{-3}$); this construction is an operational sufficiency bound rather than a proposed efficient protocol. The proposition below gives the proof. When the visibility data $\bv g,\xi,\alpha,\theta$ are calibrated, an alternative places an $h_{\mathrm{net}}$-net directly on each signed input disk at the target visibility. The output intervals then have half-width $\norm{J}_{\rm op}h_{\mathrm{net}}+e_D$ around the estimated distinguishabilities, and their union contains the complete section. This follows from the same Lipschitz argument in disk coordinates and requires only two-dimensional coverage, with the number of net points scaling as $h_{\mathrm{net}}^{-2}$ per slice. These exponents concern geometric covering numbers rather than total shot counts or calibration costs. Marker-state preparation and visibility-calibration errors must be propagated, and the confidence bounds must hold simultaneously over the net. The reduction is conditional on calibrated slice geometry; it does not remove the need for trusted coverage.

\begin{proposition}[Simultaneous finite-coverage exclusion]
\label{prop:netcertificate}
Let $\{\bv x_i\}_{i=1}^N$ be an $h_{\mathrm{net}}$-net of the full input ball. Suppose one common event $\mathcal E$, of probability at least $1-\varepsilon_{\mathrm{fail}}$, guarantees $|\widehat V_i-V(\bv x_i)|\le e_V$ and $|\widehat D_i-D(\bv x_i)|\le e_D$ for all $i$. For a target visibility band $\mathcal J$, put
\begin{align}
 a_V&=L_Vh_{\mathrm{net}}+e_V,\qquad a_D=L_Dh_{\mathrm{net}}+e_D,\nonumber\\
 \mathcal I_{\mathcal J}&=\{i:[\widehat V_i-a_V,\widehat V_i+a_V]
                  \cap\mathcal J\ne\varnothing\},\nonumber\\
 \mathcal O_{\mathcal J}&=\bigcup_{i\in\mathcal I_{\mathcal J}}
       [\widehat D_i-a_D,\widehat D_i+a_D].
 \label{eq:bandouter}
\end{align}
On $\mathcal E$, every attainable pair with $V\in\mathcal J$ has $D\in\mathcal O_{\mathcal J}$. Thus any open interval disjoint from $\mathcal O_{\mathcal J}$ is excluded throughout that visibility band.
\end{proposition}
\begin{proof}
For each input choose a net point within distance $h_{\mathrm{net}}$. The Lipschitz bounds and $\mathcal E$ place its pair inside the rectangle of half-widths $(a_V,a_D)$ centered at that net point's estimates. If its visibility lies in $\mathcal J$, that rectangle is indexed by $\mathcal I_{\mathcal J}$, proving the containment.
\end{proof}
The proposition certifies an exclusion, including for visibility bands that may be empty of physical points. To certify a disconnected section, one must also establish two attainable values on opposite sides of the excluded interval at a common exact visibility. Attainable points at different visibilities within a band do not suffice. The common confidence event must cover calibration, estimation, and any choice of axis or time made from the same data; independent validation data provide an alternative.

For the static channel at $V=0.2$, calibrated signed disks permit a simple conditional certificate. A polar net with covering radius $h_{\mathrm{net}}=0.01$ on each disk has disk-coordinate distinguishability bound $\norm{J}_{\rm op}=\epsilon=0.03$. If all reported distinguishabilities have simultaneous error at most $e_D=0.005$, each sampled value generates an outer interval of radius $0.0053$. Evaluating the ideal model on this net leaves an excluded interval approximately $(0.22982,0.35218)$. The two disk centers, $\bv x=(0,0,\pm0.2/b_0)$, are physical inputs at exactly $V=0.2$ \emph{in the calibrated theoretical model}, with $D\simeq0.19818$ and $0.38382$ on opposite sides. A possible measurement sequence is to prepare these two inputs, determine the real and imaginary fringe components, and estimate the conditional marker Bloch vectors for the optimal-discrimination calculation; a signed-disk net with common error bounds is then needed to exclude every intervening value. The two center measurements alone do not certify a gap. This calculation assumes the common error event in Proposition~\ref{prop:netcertificate}. Preparation and visibility errors can be included through the band construction above. The verification code records the net and interval union.

\subsection{Finite-shot measurement requirements for a calibrated input}
\label{sec:shotrequirements}
The finite-net constructions above assume simultaneous estimates of $V$ and $D$. A simple measurement-level bound illustrates what that requirement entails. Consider the ideal experiment before post-interaction quanton noise, with balanced priors, a calibrated fringe-phase reference, and independent preparations. For one fixed marker input, let $\widehat{\bv r}_0$ and $\widehat{\bv r}_1$ be the path-conditioned Bloch-vector estimates obtained from $n_D$ independent binary shots for each of the three Pauli observables on each conditional output state. Then $D=\norm{\bv r_0-\bv r_1}/2$. Hoeffding's inequality \citep{SUPPHoeffding1963} and a union bound over the six estimated coordinates give
\begin{equation}
 \Pr\{ |\widehat D-D|>e_D\}
 \leq 12\exp(-n_D e_D^2/6).
 \label{eq:shotsD}
\end{equation}
Indeed, every estimated Pauli expectation differs from its mean by at most $q$ except with probability $2e^{-n_Dq^2/2}$. On their common event, each reconstructed Bloch vector has Euclidean error at most $\sqrt3 q$, so the reverse triangle inequality bounds the distinguishability error by $\sqrt3 q$. Equation~\eqref{eq:shotsD} follows upon setting $q=e_D/\sqrt3$. The derivation assumes ideal, labelled access to the two conditional states; state-preparation errors and imperfect readout must be incorporated separately.

For the same input, obtain the real and imaginary components of the complex fringe amplitude as binary path-interference expectations, with $n_V$ independent shots per component and an ideal calibrated phase reference. Visibility is their Euclidean norm, so
\begin{equation}
 \Pr\{ |\widehat V-V|>e_V\}
 \leq 4\exp(-n_V e_V^2/4).
 \label{eq:shotsV}
\end{equation}
For a net containing $N$ inputs, both errors hold simultaneously with failure probability at most $\varepsilon_{\mathrm{fail}}$ if
\begin{equation}
 \begin{split}
 n_D&\geq \frac{6}{e_D^2}\ln\!\left(\frac{24N}{\varepsilon_{\mathrm{fail}}}\right),\\
 n_V&\geq \frac{4}{e_V^2}\ln\!\left(\frac{8N}{\varepsilon_{\mathrm{fail}}}\right).
 \end{split}
 \label{eq:shotsresource}
\end{equation}
This allocates failure probability $\varepsilon_{\mathrm{fail}}/(2N)$ to each type of estimate at each input; it does not assume independence between the two estimators. For one input, $e_D=0.02$ and a $5\%$ distinguishability-failure allowance alone, Eq.~\eqref{eq:shotsD} requires $n_D\geq\lceil 6(0.02)^{-2}\ln(12/0.05)\rceil=82{,}210$ shots \emph{per Pauli observable per conditional state}, or $493{,}260$ conditional-state shots for the six observables, before visibility measurements, readout, preparation, or conditional-postselection overhead. Covering multiple input settings and imposing a common confidence event increases this requirement. These conservative shot counts cover statistical estimation only. Applying Proposition~\ref{prop:netcertificate} also requires calibrated bounds on preparation, geometry, readout, and drift.

\section{Discrimination tasks and path noise}\label{supp:labels}
\subsection{Derivation of the duality bound}\label{supp:duality}
The two operational measures obey the established wave--particle
duality relation in this model. Let $|\psi_\Phi\rangle$ purify
$\rho_\Phi$. The overlap between the two conditional purified
marker states is
$\langle\psi_\Phi|(W\otimes\mathbb{I}_2)|\psi_\Phi\rangle
=\Tr(\rho_\Phi W)=\Gamma$.
Discarding the purifying system and applying the same marker
channel $\Psi$ to both paths cannot decrease fidelity. Hence
\begin{equation}
 F(\rho_0,\rho_1)\geq |\Tr(\rho_\Phi W)|.
 \label{eq:fidelitytagging}
\end{equation}
The weighted trace-distance bound and $\alpha^2\leq4p(1-p)$ therefore give
\begin{align}
 D^2&\leq 1-4p(1-p)F(\rho_0,\rho_1)^2\nonumber\\
 &\leq 1-\alpha^2|\Tr(\rho_\Phi W)|^2=1-V^2,
 \label{eq:fidelityduality}
\end{align}
The last inequality proves main-text Eq.~\eqref{eq:standardduality}.

For a pure prepared marker and $\Psi=\mathrm{id}$, the factor $\Tr(\rho_\Phi W)$ is the overlap of the two conditional pure states. For a balanced pure path source, its modulus is the fringe visibility. For mixed markers, Eq.~\eqref{eq:fidelitytagging} supplies the inequality needed here.
\subsection{Unambiguous discrimination of noisy markers}
\label{sec:usd}
Unambiguous state discrimination (USD) allows an inconclusive outcome but forbids incorrect conclusive identifications. For nonzero priors $p$ and $1-p$, a three-outcome positive operator-valued measure (POVM) $\{E_0,E_1,E_?\}$ obeys $E_j\succeq0$, $\sum_jE_j=\mathbb{I}_2$, and
\begin{equation}
 \begin{aligned}
 \Tr(E_0\rho_1)&=\Tr(E_1\rho_0)=0,\\
 P_{\mathrm{USD}}&=p\Tr(E_0\rho_0)\\
 &\quad +(1-p)\Tr(E_1\rho_1).
 \end{aligned}
 \label{eq:usddef}
\end{equation}
The general support criterion for mixed-state USD is known \citep{SUPPRudolph2003}. The support structure of mixed marker states motivates the minimum-error discrimination task used here.
\begin{proposition}[A full-rank prepared marker state precludes unambiguous which-path identification]
\label{prop:usd}
In Eq.~\eqref{eq:read}, if the prepared qubit marker has $\norm{\bv r}<1$, then $P_{\mathrm{USD}}^{\mathrm{opt}}=0$ for every pair of path-dependent marker unitaries and every common CPTP post-interaction marker channel $\Psi$. This statement does not force the minimum-error distinguishability $D$ to vanish.
\end{proposition}
\begin{proof}
Since $\rho_\Phi$ is positive definite, there are constants $0<m\leq M$ such that $m\mathbb{I}_2\preceq U_j\rho_\Phi U_j^\dagger\preceq M\mathbb{I}_2$ for both paths. Positivity of $\Psi$ gives $m\Psi(\mathbb{I}_2)\preceq\rho_j\preceq M\Psi(\mathbb{I}_2)$; hence both output states have the same support. An effect with zero probability on $\rho_1$ has zero probability on $\rho_0$, and conversely. Thus the two conclusive success terms in Eq.~\eqref{eq:usddef} vanish for every admissible USD measurement.
\end{proof}
For example, take balanced paths, $\Psi=\mathrm{id}$, $U_1=\mathbb{I}_2$, $U_0=\exp(-i\,\theta\sigma_x/2)$, and a marker initially in $|0\rangle$. Depolarizing preparation produces $\rho_\Phi=(1-\epsilon)|0\rangle\langle0|+\epsilon\mathbb{I}_2/2$. For $0<\theta\leq\pi$,
\begin{equation}
 \begin{aligned}
 D(\epsilon)&=(1-\epsilon)|\sin(\theta/2)|,\\
 P_{\mathrm{USD}}^{\mathrm{opt}}(\epsilon)
 &=\begin{cases}
 1-|\cos(\theta/2)|,&\epsilon=0,\\
 0,&0<\epsilon\leq1.
 \end{cases}
 \end{aligned}
 \label{eq:usdnoise}
\end{equation}
The first USD value is the standard equal-prior pure-state result; the abrupt loss of unambiguous success for any positive depolarization reflects the zero-error constraint; the Helstrom distinguishability remains continuous. Pure prepared states ($\norm{\bv r}=1$) permit nonzero USD only if the conditional output supports allow it; this purity condition is necessary here, not sufficient. Imperfect implementation of a nominal USD POVM may also generate incorrect conclusive outcomes and is no longer an error-free measurement. The regions studied here use the continuously varying minimum-error distinguishability. Schemes permitting both errors and inconclusive outcomes have been treated in complementarity settings \citep{SUPPBagan2020}; fixed-error-margin discrimination is a separate problem \citep{SUPPSugimoto2009}.

\subsection{Path noise and the inferred label}
\label{sec:postquanton}
We now consider a phase-covariant path channel acting locally on the quanton after the which-path interaction. Its transverse Bloch components are rotated about $z$ and multiplied by $\kappa_Q\geq0$. Write $Z_Q=|0\rangle\langle0|-|1\rangle\langle1|$ for the path-population observable, so the source imbalance is $\delta=\Tr(\rho_Q Z_Q)$. The channel acts on the longitudinal Bloch component as
\begin{equation}
 \delta\longmapsto\lambda_Q\delta+a_Q.
 \label{eq:pathchannel}
\end{equation}
All parameters are constrained by CPTP physicality. This family includes depolarizing noise, phase damping, and generalized amplitude damping; for the last, $\kappa_Q=\sqrt{1-\gamma_Q}$, $\lambda_Q=1-\gamma_Q$, and $a_Q=\gamma_Q(2q_Q-1)$, where $\gamma_Q\in[0,1]$ is the damping strength and $q_Q\in[0,1]$ is the equilibrium population of $|0\rangle$. An additional phase-damping factor may multiply $\kappa_Q$.

Let $V$ denote Eq.~\eqref{eq:V}, the visibility before this path channel. The measured fringe visibility is
\begin{equation}
 V_{\mathrm{obs}}=\kappa_Q V.
 \label{eq:Vobs}
\end{equation}
The marker ensembles conditioned on the original path are unchanged by a local trace-preserving path channel. Thus the original-path distinguishability $D$ in Eq.~\eqref{eq:D} remains a well-defined discrimination task. It can differ from the optimal distinguishability $D_{\mathrm{out}}$ for inferring the final path population from the marker after noise. Define
\begin{equation}
 K_+=pR_0+(1-p)R_1,\qquad \delta_{\mathrm{out}}=\lambda_Q\delta+a_Q.
 \label{eq:Hpf}
\end{equation}
The final-path distinguishability is
\begin{equation}
 D_{\mathrm{out}}=\max\left\{|\delta_{\mathrm{out}}|,
 \norm{B(\lambda_QK_-+a_QK_+)\bv r+\delta_{\mathrm{out}}\bv h}\right\}.
 \label{eq:Dfinal}
\end{equation}
To see this, the unnormalized marker blocks associated with final $Z$ outcomes have difference
\begin{equation}
 \Delta_{\mathrm{out}}=\lambda_Q[p\rho_0-(1-p)\rho_1]
             +a_Q[p\rho_0+(1-p)\rho_1].
 \label{eq:deltafinal}
\end{equation}
It follows by applying $\Lambda_Q^\dagger(Z_Q)=\lambda_QZ_Q+a_Q\mathbb{I}_2$ to the joint state; off-diagonal path blocks do not contribute. Taking its trace norm gives Eq.~\eqref{eq:Dfinal}. The source coherence affects $V_{\mathrm{obs}}$, but not this population-inference formula.

Theorem~\ref{thm:region} therefore applies to both tasks. For $(V_{\mathrm{obs}},D)$ replace $\alpha$ by $\alpha_{\mathrm{eff}}=\kappa_Q\alpha$. For $(V_{\mathrm{obs}},D_{\mathrm{out}})$ also replace the quadratic data and prior predictability by
\begin{align}
 A_{\mathrm{out}}&=B(\lambda_QK_-+a_QK_+)T,\nonumber\\
 \bv d_{\mathrm{out}}&=B(\lambda_QK_-+a_QK_+)\bv t+\delta_{\mathrm{out}}\bv h,
 \qquad P_{\mathrm{out}}=|\delta_{\mathrm{out}}|.
 \label{eq:finaldata}
\end{align}
The input disks and their quadratic proof are otherwise unchanged; if $\kappa_Q=0$, the observed visibility is zero and the full-ball case applies. The replacements are algebraic and do not require reinterpreting $\delta_{\mathrm{out}}$ as the original source prior.

If the path channel is unital ($a_Q=0$), then $D_{\mathrm{out}}=|\lambda_Q|D$. The balanced-source nonunital-preparation gap witness consequently carries over to final-path inference whenever $\kappa_Q>0$ and $\lambda_Q\neq0$. For a nonunital path channel, the $a_QK_+$ term and channel offset can break this symmetry. A final-path gap then need not witness nonunital marker preparation; Sec.~\ref{app:finalgap} gives a physical counterexample. The original-path gap witness remains valid at a resolved $V_{\mathrm{obs}}$ when $\kappa_Q>0$. Noise during the marker interaction, joint path--marker noise, and arbitrary path channels mixing longitudinal and transverse components require a different region analysis.

\subsection{A final-path gap with unital marker preparation}
\label{app:finalgap}
Consider a unital measure-and-prepare marker channel with
\begin{equation}
 T=(0.6,0,0.6)^{\mathsf T}\bv e_z^{\mathsf T},\qquad \bv t=0.
\end{equation}
It measures $\sigma_z$ on the marker and prepares states with opposite Bloch vectors $\pm(0.6,0,0.6)$, whose norms are below one. Use a balanced pure path source, $U_0=-i\,\sigma_z$, $U_1=\mathbb{I}_2$, amplitude damping of strength $1/2$ on the marker after the marker interaction, and path-qubit amplitude damping of strength $1/5$ after the marker interaction. The post-interaction marker channel has $B=\operatorname{diag}(1/\sqrt2,1/\sqrt2,1/2)$ and $\bv h=\bv e_z/2$; the path map has $\kappa_Q=\sqrt{0.8}$, $\lambda_Q=0.8$, and $a_Q=0.2$. At
\begin{equation}
 V_{\mathrm{obs}}=\sqrt{0.8}\,(0.54)=0.482991\ldots,
\end{equation}
the fixed-visibility input slices have $x_z=\pm0.9$. Since $T$ depends only on $x_z$, each slice gives one final-path distinguishability:
\begin{equation}
 D_{\mathrm{out}}=\sqrt{(0.8)^2(0.54)^2/2+(0.1\pm0.054)^2}.
\end{equation}
The complete section is consequently the two-point set $\{0.308914\ldots,0.342094\ldots\}$, which is disconnected even though marker preparation is unital. In contrast, the original-path distinguishability is the same on both slices, $D=0.54/\sqrt2$. This counterexample isolates the change in the inference task caused by a nonunital path channel.

\section{Lindblad realization and numerical methods}\label{supp:dynamics}
\subsection{Time-independent Lindblad generator for the gap family}
\label{app:gksl}
Setting $\hbar=1$, we use the Gorini--Kossakowski--Sudarshan--Lindblad (GKSL) convention
\begin{equation}
 \dot\rho=-i\,[H,\rho]+\sum_{j,k=1}^3
 \mathsf K_{jk}\left(\sigma_j\rho\sigma_k-\tfrac12\{\sigma_k\sigma_j,\rho\}\right),
 \label{eq:generatorGKSL}
\end{equation}
where the positive semidefinite Hermitian matrix $\mathsf K$ is represented in the Pauli basis $(\sigma_x,\sigma_y,\sigma_z)$. A Hamiltonian and matrix producing Eq.~\eqref{eq:markovbloch} are
\begin{equation}
 H=\frac{\chi(\gamma_x-\gamma_z)}4\sigma_y,
 \label{eq:markovH}
\end{equation}
\begin{equation}
 \mathsf K=\frac14\begin{pmatrix}
 \gamma_z&0&\chi(\gamma_x-\gamma_z)\\
 0&\gamma_z&-i\,t_0\gamma_x\\
 \chi(\gamma_x-\gamma_z)&i\,t_0\gamma_x&2\gamma_x-\gamma_z
 \end{pmatrix}.
 \label{eq:markovK}
\end{equation}
The first two diagonal pivots are positive, and the remaining Schur complement is
\begin{equation}
 s_K=\frac{2\gamma_x-\gamma_z}4-
 \frac{\chi^2(\gamma_x-\gamma_z)^2+t_0^2\gamma_x^2}{4\gamma_z}.
 \label{eq:markovSchur}
\end{equation}
Thus $\mathsf K\succeq0$ if and only if Eq.~\eqref{eq:markovphysical} holds for this representation. Equivalently, with $\eta=\gamma_z/\gamma_x\in(0,1)$, the admissibility condition is $t_0^2+(1+\chi^2)(1-\eta)^2\leq1$. Writing $L_j=\bv l_j\cdot\bv\sigma$, one set of jump operators satisfying $\mathsf K=\sum_j\bv l_j\bv l_j^\dagger$ is
\begin{align}
 L_1&=\frac{\sqrt{\gamma_z}}2\sigma_x+
 \frac{\chi(\gamma_x-\gamma_z)}{2\sqrt{\gamma_z}}\sigma_z,\nonumber\\
 L_2&=\frac{\sqrt{\gamma_z}}2\sigma_y+
 \frac{i\,t_0\gamma_x}{2\sqrt{\gamma_z}}\sigma_z,\nonumber\\
 L_3&=\sqrt{s_K}\,\sigma_z.
 \label{eq:markovjumps}
\end{align}
Inserting $L_j$ into $\sum_j[L_j\rho L_j^\dagger-\{L_j^\dagger L_j,\rho\}/2]$
reproduces Eq.~\eqref{eq:generatorGKSL}. The dissipator generates
an affine translation rate $t_0\gamma_x\bv e_x$ and a symmetric
coupling between $r_x$ and $r_z$. The Hamiltonian cancels the
contribution from $r_x$ to $\dot r_z$ while reinforcing the
contribution from $r_z$ to $\dot r_x$, giving
Eq.~\eqref{eq:markovbloch}. This is an engineered, time-independent
open-system generator rather than a claim of a unique microscopic bath.

\subsection{Optimal time for every Lindblad visibility slice}
\label{app:generaloptimum}
For Eq.~\eqref{eq:markovwidth}, fix $0<\zeta<1$, $q=\sqrt{1-\zeta^2}$, and $k=\chi\zeta>0$, with $t_0>0$. Half the unclipped gap is the smaller of
\begin{align}
 F_{\mathrm d}(\tau)&=t_0-(t_0+q)e^{-\gamma_x\tau},\\
 F_{\mathrm c}(\tau)&=ke^{-\gamma_z\tau}-(k+q)e^{-\gamma_x\tau}.
\end{align}
The contribution $F_{\mathrm d}$ from displacement is strictly increasing. The contribution $F_{\mathrm c}$ from longitudinal--transverse coupling has a unique maximum at
\begin{equation}
 \tau_{\mathrm{peak}}=\frac{\ln[\gamma_x(k+q)/(\gamma_zk)]}{\gamma_x-\gamma_z}.
 \label{eq:generaltime}
\end{equation}
Write $\tau_{\mathrm{opt}}=\tau_{\mathrm{opt}}(\zeta)$ for the time maximizing the gap width at the fixed fractional visibility. If $F_{\mathrm d}(\tau_{\mathrm{peak}})\geq F_{\mathrm c}(\tau_{\mathrm{peak}})$, then $\tau_{\mathrm{opt}}=\tau_{\mathrm{peak}}$. Otherwise $\tau_{\mathrm{opt}}>\tau_{\mathrm{peak}}$ is the unique later solution of
\begin{equation}
 t_0(1-e^{-\gamma_x\tau_{\mathrm{opt}}})
 =k(e^{-\gamma_z\tau_{\mathrm{opt}}}-e^{-\gamma_x\tau_{\mathrm{opt}}}).
 \label{eq:generalcrossing}
\end{equation}
To verify uniqueness, set $a=e^{-\gamma_x\tau}$ and $b=e^{-\gamma_z\tau}$. The ratio $(b-a)/(1-a)$ decreases strictly from $1-\gamma_z/\gamma_x$ to zero. Indeed it is $1-(1-e^{-\gamma_z\tau})/(1-e^{-\gamma_x\tau})$; the subtracted ratio increases strictly for $0<\gamma_z<\gamma_x$, as follows by differentiating or from the secant slopes of the concave function $x^{\eta}$. Thus $F_{\mathrm d}-F_{\mathrm c}=t_0(1-a)-k(b-a)$ has at most one positive crossing, from negative to positive. Before $\tau_{\mathrm{peak}}$ both branches increase. If that crossing lies later, their minimum increases to $\tau_{\mathrm{opt}}$ and then follows the decreasing $F_{\mathrm c}$; otherwise its maximum occurs at $\tau_{\mathrm{peak}}$.

For the high-visibility example in main-text Eq.~\eqref{eq:highvisibilityparameters}, $a=1/100$, $b=4/5$, $t_0=1/4$, $\chi=1/8$, and $\zeta=3/4$. The two displacement contributions satisfy $t_0(1-a)=99/400$ and $\chi(b-a)\zeta=237/3200$, while $a\sqrt{1-\zeta^2}=\sqrt7/400$. Consequently the coupling contribution limits the width, yielding exactly $\Delta D=(237-8\sqrt7)/1600$ as quoted in the main text.

For the main-text rates and $\zeta=1/2$, the coupling term is the active one throughout. The stationary-point identity $\gamma_x(1+2\sqrt3)a=\gamma_z b$ yields $a/b=\eta/(1+2\sqrt3)$ and hence $b=[\eta/(1+2\sqrt3)]^{\eta/(1-\eta)}$, giving the expressions in main-text Eq.~\eqref{eq:markovmaxexact}. These identities concern a fixed \emph{fraction} of the time-dependent maximum visibility, not a fixed absolute visibility.

Thus the maximizing time is either the maximum of the coupling contribution or the unique later crossing of the two branches, found from one scalar equation. The crossing need not have an elementary closed-form expression for arbitrary rates. For $\gamma_x=2\gamma_z$ in the crossing case, putting $y=e^{-\gamma_z\tau_{\mathrm{opt}}}$ gives $t_0(1+y)=ky$ and hence $y=t_0/(k-t_0)$; a positive crossing requires $k>2t_0$.

\subsection{Optimization over visibility fraction and time}
\label{app:globaloptimum}
For the same Lindblad family, consider in addition the closed
range $0\leq\zeta\leq1$. With
$a=e^{-\gamma_x\tau}$, $b=e^{-\gamma_z\tau}$,
$d=t_0(1-a)$ and $h=\chi(b-a)$, Eq.~\eqref{eq:markovwidth}
gives
\begin{equation}
 \Delta D(\tau,\zeta)=2\left[
 \min(d,h\zeta)-a\sqrt{1-\zeta^2}
 \right]_+.
 \label{eq:globalgap}
\end{equation}
Both contributions inside the square brackets are
nondecreasing in $\zeta$ for fixed $\tau$; hence
\begin{equation}
 \max_{0\leq\zeta\leq1}\Delta D(\tau,\zeta)
 =2\min\{d,h\}.
 \label{eq:globalfraction}
\end{equation}
Assume $t_0>0$, $\chi>0$ and $0<\gamma_z<\gamma_x$.
The function $d(\tau)$ is increasing while $h(\tau)$
is maximal at
$\tau_p=\ln(\gamma_x/\gamma_z)/(\gamma_x-\gamma_z)$.
If $d(\tau_p)\geq h(\tau_p)$, then the global optimum is
$\tau_p$. With $\eta=\gamma_z/\gamma_x$,
\begin{equation}
 \begin{split}
 V_{\rm global}&=\eta^{\eta/(1-\eta)},\\
 \Delta D_{\rm global}&=
 2\chi(1-\eta)\eta^{\eta/(1-\eta)}.
 \end{split}
 \label{eq:globalpeak}
\end{equation}
If instead $d(\tau_p)<h(\tau_p)$, the optimum occurs
at the unique later solution of $d=h$. The monotonic ratio
$(b-a)/(1-a)$ used in Sec.~\ref{app:generaloptimum}
proves uniqueness; the minimum of $d$ and $h$ increases
up to that crossing and decreases afterwards. When
$t_0\chi=0$ the gap is identically zero.

For main-text Eq.~\eqref{eq:markovparameters}, the first
case applies. The optimum has
$\tau_p/\tau_0=0.5281460669\ldots$,
$V_{\rm global}=0.6308494709\ldots$ and
$\Delta D_{\rm global}=0.4739225042\ldots$.
The two endpoint distinguishabilities are
$0.0159606579\ldots$ and $0.4898831621\ldots$.
At $\zeta=1$ both signed disks have zero radius,
so these are two attainable points, not separated
positive-width intervals. Restricting to
$0<\zeta<1$ makes the quoted global value a
supremum rather than an attained maximum. In contrast,
the main-text $\zeta=1/2$ optimum retains two
positive-width intervals and remains unchanged.

\subsection{Evolution at fixed absolute visibility}
\label{app:fixedabsolute}
The preceding optimizations hold the visibility fraction fixed
or vary it together with time. Now fix an absolute visibility
$0<V<1$ and assume $t_0\chi>0$ and $\gamma_x>\gamma_z>0$,
with the complete-positivity condition in
Eq.~\eqref{eq:markovphysical}. This visibility is attainable
for $0\leq\tau\leq\tau_V=\ln(1/V)/\gamma_z$.
Put $\eta=\gamma_z/\gamma_x\in(0,1)$ and $b=e^{-\gamma_z\tau}$,
so $V\leq b\leq1$ and $a=b^{1/\eta}$. Substituting
$\zeta=V/b$ into Eq.~\eqref{eq:markovwidth} gives
\begin{equation}
 \begin{split}
 \Delta D_V(\tau)=2\Big[&
 \min\{t_0(1-b^{1/\eta}),\chi V(1-b^{1/\eta-1})\}\\
 &-b^{1/\eta-1}\sqrt{b^2-V^2}\Big]_+.
 \end{split}
 \label{eq:fixedabsolutewidth}
\end{equation}
As time increases, $b$ decreases. Both entries of the minimum
increase strictly, while the final term decreases strictly
for $V<b<1$. The unclipped expression is therefore strictly
increasing. It starts negative and ends positive, so there is
one onset time; the gap then increases strictly up to $\tau_V$.
Its maximum is
\begin{equation}
 \max_{0\leq\tau\leq\tau_V}\Delta D_V(\tau)
 =2\min\{t_0(1-V^{1/\eta}),\chi V(1-V^{1/\eta-1})\}.
 \label{eq:fixedabsolutemax}
\end{equation}
At $\tau_V$, each signed input disk is a point. If both disks
are required to have positive radius, this value is a supremum,
approached as $\tau$ approaches $\tau_V$ from below. When $t_0\chi=0$, the
gap is identically zero. For the main-text parameters and
$V=0.2$, the maximum is $0.19845073\ldots$ at
$\tau_V/\tau_0=1.84510623\ldots$.

\subsection{Numerical verification}
\label{sec:verification}
The plotted boundaries and dynamical curves were calculated from the
analytical expressions. As an independent check, we propagated marker density matrices
through randomly generated CPTP preparation and subsequent marker
channels, applied the two path-dependent marker unitaries, and evaluated fringe
visibility and Helstrom distinguishability directly. We compared
these values with the Bloch-vector formulas. For 600 randomly generated
physical configurations, four-Kraus channels were obtained by
orthonormalizing complex Gaussian matrices; marker inputs were
sampled uniformly in Bloch-ball volume, with varied path priors
and path-dependent marker rotations. Direct calculations agreed to within
$1.2\times 10^{-15}$. Fixed-visibility endpoints were checked against
direct disk optimization. For 180 randomly drawn quadratic disk problems,
constrained minimization used sequential least squares programming
(SLSQP) with a function tolerance of
$2\times 10^{-13}$ and at most 500 iterations; maxima were independently
bounded using a $16\,384$-point angular grid together with an
analytical curvature bound on the angular discretization error.
The largest absolute
difference for the independent minima was $4.5\times10^{-9}$.
\mbox{Singular} forms and collapsed disks were also tested against the
scalar-equation construction in Sec.~\ref{app:trust}.

For the dynamical model, the analytical Bloch evolution was checked against
matrix exponentiation of the Lindblad generator and numerical integration
of the master equation (relative and absolute integration tolerances
$2\times10^{-11}$ and $2\times10^{-13}$).
The integration comparison agreed at the $10^{-12}$ level.
We also tested the fixed-absolute-visibility behavior, including the
monotonicity and gap widths at the last time for which the prescribed
visibility remains attainable, as derived in Sec.~\ref{app:fixedabsolute}.
At the terminal time the condition $b=V$ was imposed exactly to avoid an
artificial disk radius caused by roundoff. Sampling details, random seeds,
tolerances, and individual discrepancies are documented with the
verification code, available from the corresponding author upon
reasonable request. The finite-coverage examples assume calibration
and simultaneous error bounds; they do not represent experimental data.

\begingroup
\let\origbiblabel\label
\renewcommand{\label}[1]{%
  \ifnum\pdfstrcmp{#1}{LastBibItem}=0
    \origbiblabel{SupplementLastBibItem}%
  \else
    \origbiblabel{#1}%
  \fi}

\endgroup

\end{document}